\documentclass[preprint,12pt,authoryear]{elsarticle}

\usepackage{amssymb}

\usepackage{lscape}
\usepackage{float}
\usepackage{graphicx} 

\usepackage{longtable}
\usepackage{xcolor}
\usepackage{siunitx}
\usepackage{eqnarray,amsmath}

\journal{Physics of the Earth and Planetary Interiors}

\begin{document}

\begin{frontmatter}




\title{From Geostrophic to Magnetically-Damped Turbulence in Liquid Metal Rotating Magnetoconvection}



\author[inst1]{Tao Liu}
\author[inst2]{Yufan Xu}
\author[inst1]{Jewel A. Abbate}
\author[inst3]{Jonathan S. Cheng}
\author[inst1]{Jonathan M. Aurnou}

\affiliation[inst1]{organization={Earth, Planetary, and Space Sciences, University of California},
            city={Los Angeles},
            state={CA},
            country={USA}}

\affiliation[inst2]{organization={Princeton Plasma Physics Laboratory},
            city={Princeton},
            state={NJ},
            country={USA}}

\affiliation[inst3]{organization={Mechanical and Nuclear Engineering,  United States Naval Academy},
            city={Annapolis},
            state={MD},
            country={USA}}

\begin{abstract}

Understanding planetary core convection dynamics requires the study of convective flows in which the Coriolis and Lorentz forces attain a leading-order, so-called magnetostrophic balance. 
Experimental investigations of rotating magnetoconvection (RMC) in the magnetostrophic regime are therefore essential to broadly characterize the properties of local-scale planetary core flow. Towards this end, we present here the first thermovelocimetric measurements of magnetostrophic, liquid metal convection,
which are made using liquid gallium as the working fluid, at moderate rotation rates (Ekman numbers $10^{-4} \leq Ek\leq 10^{-5}$) and in the presence of dynamically strong magnetic fields (Elsasser number $\Lambda=1$). Complementary rotating convection (RC) experiments are performed at the same rotation rates to serve as reference cases. 
Our RMC velocity measurements adequately follow a geostrophic turbulent scaling for cases in which local-scale convective inertial forces exceed the Lorentz forces in the fluid bulk.  In cases where Lorentz forces exceed local-scale inertia ($N_\ell \gtrsim 3$), the root-mean-square RMC velocities are magnetically damped, yielding values below the geostrophic turbulent RC scaling prediction. An enhancement in heat transfer is observed, which we attribute to the increased coherence of vertically aligned magnetostrophic convective flow.
Extrapolating these laboratory results, we predict that convection-scale flows in Earth's core occur in the magnetically damped $N_\ell \gtrsim 3$ regime with Rayleigh number values between $10^{24}$ and $10^{26}$. 

\end{abstract}



\begin{keyword}
Core dynamics\sep Magnetostrophy \sep Convective turbulence 



\end{keyword}

\end{frontmatter}


\section{Introduction}
\label{intro}

Earth’s outer core dynamics are primarily governed by 
convective turbulent flows of molten iron and nickel, which sustain the geodynamo process responsible for generating the planet’s magnetic field \citep{Roberts_2013}. These complex fluid motions are driven by buoyancy-induced convection, influenced by rapid rotation and electromagnetic forces within the liquid outer core \citep{Jones_2011}. Under the influence of the Coriolis effect induced by Earth's rapid rotation, outer core flows tend to organize into anisotropic, columnar modes that are elongated parallel to the rotation axis, a phenomenon known as the Proudman–Taylor (P-T) constraint \citep{Proudman1916, Taylor1917,Greenspan1969}. However, linear theory suggests that when the Coriolis and Lorentz forces reach a leading-order ``magnetostrophic'' balance, the P-T constraint is relaxed.  It has long been postulated that this magnetostrophic relaxation allows planetary core convection to develop into an optimal state, facilitating more efficient heat and compositional transport within the outer core \citep{ChandrasekharElbert1955, Eltayeb1972, KingAurnou2015PNAS, yadav2016effect, HornAurnou2025}. However, few studies have characterized the properties of magnetostrophic convective flow. Thus, the properties of magnetostrophic convection are not well known. Further, it is not known if magnetostrophic convection is indeed optimal for generating dynamo action in planetary core settings. 

\begin{figure}[ht!]
  \centerline{\includegraphics[width=15cm]{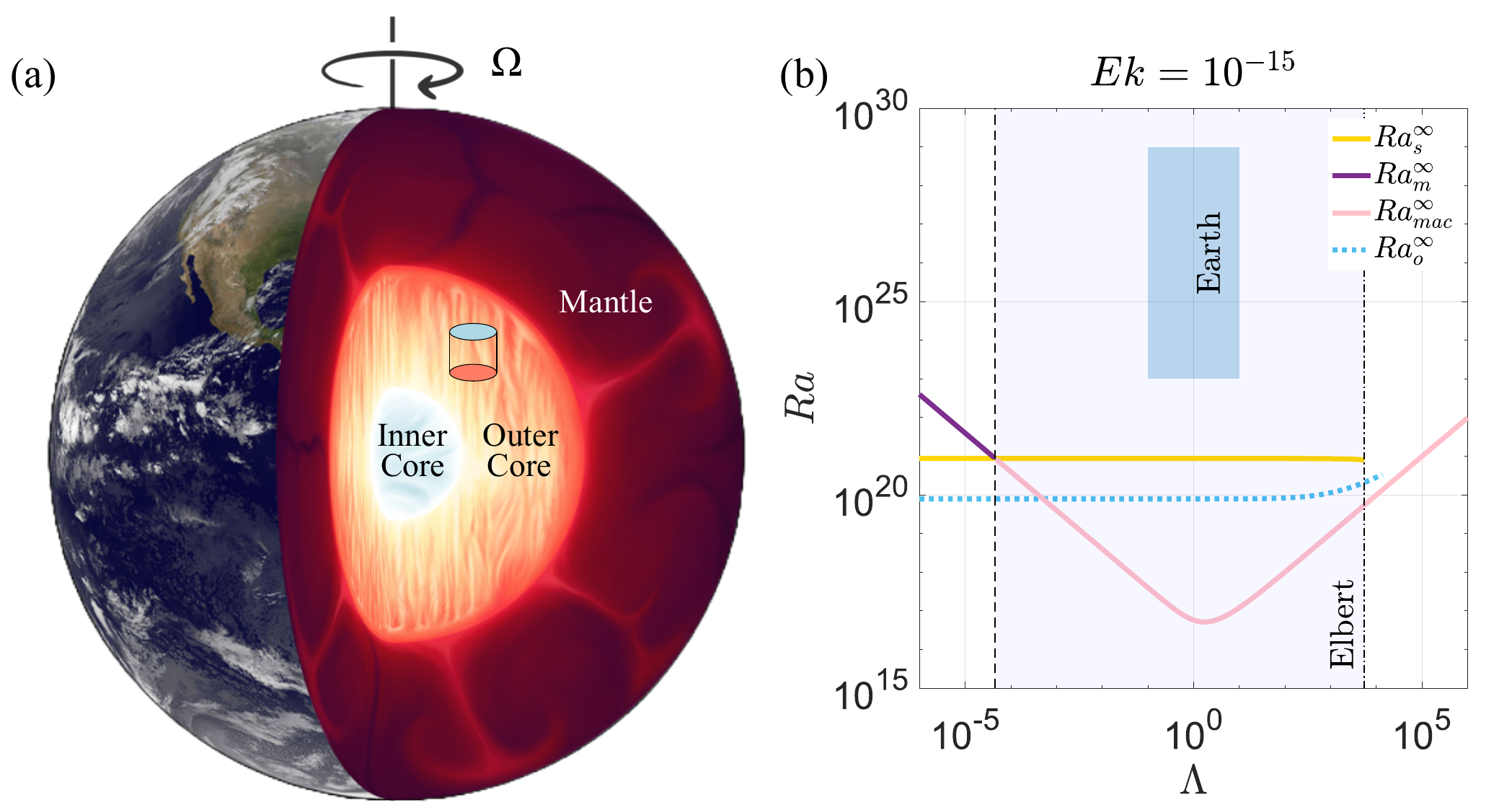}}
  \caption{(a) Illustration of earth interior with a cylindrical container in outer core representing the simulated laboratory setup (Earth surface image: NASA/NOAA/GOES);  (b) linear theory predictions of multi-modal rotating magnetoconvection for Earth's outer core in $Pr=0.1$ core-like fluid, semi-transparent blue shaded rectangle shows the $Ra$-$\Lambda$ range estimates for Earth's outer core from previous work \citep[e.g.,][]{Gubbins2001}.}
\label{fig:earthcore_linearprediction}
\end{figure}

Our cylindrical laboratory experiments simulate a local fluid parcel within Earth's outer core to investigate small-scale convection dynamics in electrically conducting liquid metals, as shown schematically in  Figure~\ref{fig:earthcore_linearprediction}a.
Previous experimental studies using liquid gallium have revealed evidence of magnetostrophic convective flow by demonstrating enhanced heat transfer at Elsasser number $\Lambda\sim 1$ \citep{KingAurnou2015PNAS}. However, laboratory velocity measurements of the ``optimized'' flow field claimed to develop in the magnetostrophic regime are still lacking. In this study, we employ the 'RoMag' apparatus at UCLA, filled with liquid gallium and capable of imposing both Coriolis and Lorentz forces to explore magnetostrophic dynamics in greater detail. In particular, we quantify the thermally driven flow velocities that arise in the magnetostrophic RMC regime. 

Numerical simulations in spherical shell geometries suggest that the presence of magnetic fields can enhance convective efficiency, especially in polar regions of rotating spherical shells \citep{yadav2016effect}. Further, \citet{Yadav_2016_PNAS} found a leading order geostrophic force balance in a suite of dynamo models. However, they showed that a second order magnetostrophic balance could arise on the local convective scale with the resulting flow self-organizing into slightly larger-scale, 
axially-aligned structures, in basic agreement with the theoretical arguments of \cite{calkins2015multiscale} and \cite{aurnou2017cross}. 

In this work, we focus on velocity measurements in rotating magnetoconvection (RMC) over the Ekman number range $Ek\in[10^{-4},10^{-5}]$ at the Elssaser number $\Lambda=1$, where $Ek$ describes the nondimensional rotation period and $\Lambda$ estimates the ratio of quasistatic Lorentz and Coriolis forces \citep{HornAurnou2025}. Rotating convection (RC) experiments are also conducted for comparison. The measured velocities are evaluated with respect to theoretical predictions for rotating convective turbulence, including the inviscid Coriolis–Inertial–Archimedean (CIA) balance \citep{Jones_2011} as well as the fully diffusivity-free (DF) scaling regime \citep{JulienKeith2012, abbate2024, abbate2024diffusivityfree}. The interaction parameter is also calculated to quantify the ratio of Lorentz to inertial forces. It serves as an indicator of the flow regime: a high value suggests Lorentz-force dominance characteristic of the magnetostrophic regime, while a low value indicates inertial dominance, more typical of geostrophic turbulence.

This paper is organized as follows. Section \ref{parameter_theory}  introduces the key dimensionless parameters and summarizes the theoretical predictions and scaling laws for rotating convection and rotating magnetoconvection. Section \ref{method} describes the experimental setup, measurement techniques, and explored parameter space. We present and analyze the experimental results in Section \ref{Results}, including transport scalings for both rotating convection and rotating magnetoconvection. In sections \ref{sec:Application} and \ref{Conclu}, we extrapolate our laboratory findings to Earth's outer core conditions and summarize our findings.

\section{Nondimensional parameters and theoretical background}
\label{parameter_theory}

\subsection{Nondimensional parameters}
\label{sec:nd_parameters}

Rotating convection (RC) is governed by four nondimensional parameters \citep{Horn_Schmid_2017, Kunnen04052021}. The Rayleigh number $Ra$ quantifies the effect of the competition between buoyancy and diffusive effects: 
\begin{equation}
Ra=\frac{\alpha g\Delta T H^3}{\nu\kappa},
\label{eq:Ra}
\end{equation}
where $\alpha$ is the thermal expansion coefficient; $g$ is gravitational acceleration; $\Delta T$ is the mean temperature difference across the fluid layer of depth $H$; and $\nu$ and $\kappa$ are the kinematic viscosity and the thermal diffusivity, respectively. 
The Ekman number $Ek$ characterizes the relative influence of viscous forces to Coriolis forces
\begin{equation}
Ek=\frac{\nu}{2\Omega H^2},
\label{eq:Ek}
\end{equation}
where $\Omega$ is the angular rotation rate of the system.
The Prandtl number characterizes the thermomechanical properties of the material by expressing the ratio of momentum diffusivity to thermal diffusivity:
\begin{equation}
Pr=\frac{\nu}{\kappa}.
\label{eq:Pr}
\end{equation}
In addition, the geometrical aspect ratio of the cylindrical container, 
\begin{equation}
\Gamma=\frac{D}{H},
\label{eq:Gamma}
\end{equation}
can play an important role in determining the convective flow structures and statistics \citep[e.g.,][]{weiss2011heat, cheng2018, horn2022unravelling, Xu2025}.  Here, $R$ is the radius and $D = 2R$ is the diameter of the cylinder.

A fifth nondimensional parameter is needed to describe RMC systems. In local-scale core dynamics, the Elsasser number, $\Lambda$, is typically used, as it measures the relative strength of the (quasi-static) Lorentz force to the Coriolis force \citep{Soderlund2015, dormy2016strong, aurnou2017cross, HornAurnou2025, Teed2025}: 
\begin{equation}
\Lambda=\frac{\sigma B^2}{2\rho \Omega} = Ek \, Ch,
\label{eq:Elsasser}
\end{equation}
where $\sigma$ is the electrical conductivity of the fluid, $B$ is the magnetic field strength, and $\rho$ is the density of the fluid. Alternatively, one may use the Chandrasekhar number, $Ch=\sigma B^2 H^2/(\rho \nu)$, which estimates the ratio of the Lorentz and viscous forces.

Another essential parameter is the convective Rossby number, which 
provides an \textit{a priori} estimate of the ratio of the buoyancy-driven inertial and Coriolis forces acting on the local convective scale \citep{Aurnou2020}: 
\begin{equation}
Ro_c=\sqrt{\frac{g\alpha \Delta T}{4\Omega^2H}} = \sqrt{\frac{RaEk^2}{Pr}}.
\label{eq:Roc}
\end{equation}
The convective Rossby number is often used to distinguish different convective flow regimes in rotating convection (RC) systems \citep[e.g.,][]{gastine2013zonal, gastine2013solar, mabuchi2015differential, camisassa2022solar, Soderlund2025}. 
It also scales the ratio of rapidly rotating convective flow velocities, $U_\Omega \approx \alpha g \Delta T / (2 \Omega)$, and non-rotating convective velocities, $U_{ff} \approx \sqrt{\alpha g \Delta T H}$, such that $Ro_c \sim U_\Omega/U_{ff} = Re_\Omega / Re_{ff}$ \citep{Aurnou2020}. Here, $Re = UH/\nu$ is the Reynolds number, which estimates the system-scale ratio of inertial and viscous forces; the geostrophic turbulent scaling gives $Re_\Omega \sim Ra Ek / Pr$; and the non-rotating free-fall scaling is $Re_{ff} \sim \sqrt{Ra/Pr}$. In addition, it estimates the characteristic scale, $\ell$, of geostrophic turbulent RC flows such that $\ell \approx Ro_c H$ \citep[e.g.,][]{abbate2024shadowgraph, Hadjerci_Bouillaut_Miquel_Gallet_2024}.

The local ratio of the Lorentz and inertial forces is given by $N_\ell$, the local interaction parameter, 
\begin{equation}
N_{\ell}=\frac{\sigma B^2 \ell}{\rho U}.
\label{eq:N_ell}
\end{equation}
If we assume that the convection exists in, or relatively near, the geostrophic turbulence regime, then we may take $\ell \approx Ro_c H$ and $U \approx U_\Omega$. Subbing these into (\ref{eq:N_ell}) yields what we will call the convective interaction parameter, 
\begin{equation}
N_{c}=\frac{\sigma B^2 (Ro_c H)}{\rho U_\Omega }
=\frac{Ch \, Ro_c}{Re_\Omega} \, ,
\label{eq:N_c}
\end{equation}
which can be reformulated in a variety of forms, including 
\begin{displaymath}
N_c = \frac{Ch}{Re_{ff}} = \sqrt{\frac{Ch^2 Pr}{Ra}} \quad \mbox{and} \quad N_c = \frac{\Lambda}{ Ro_c} \, .
\end{displaymath}
The similarity in structure and high degree of interconnectivity between $Ro_c = \sqrt{Ra Ek^2 / Pr}$ and $N_c^{-1} = \sqrt{Ra Ch^{-2} / Pr}$ arise here because we are using the quasi-static form of the Lorentz force density $f_L \sim \sigma UB^2$, as applies in laboratory experiments \citep{HornAurnou2025}, and also because we are choosing geostrophic turbulence estimates for the characteristic scales of $\ell$ and $U$. In contrast, the interaction parameter is not nearly so similar in structure to $1/Ro_c$ in the ageostrophic, non-quasistatic dynamo modeling study of \cite{Soderlund2025}. 





\subsection{Convection Modes}
\label{rotating convection}


In rotating convection in a cylinder, convection can occur via three modes of instability: oscillatory thermal-inertial, steady geostrophic, and sidewall-attached modes. We briefly introduce here the theoretical predictions for the onset and the frequency of each mode.


The oscillatory bulk mode typically dominates near onset in low-Prandtl-number fluids and is characterized by time-periodic, columnar flows spanning the fluid layer \citep{Chandrasekhar1961, JULIEN_KNOBLOCH_1998, Julien1999Book, ZHANG_LIAO_2009, Vogt_Horn_Aurnou_2021, abbate2024diffusivityfree, Xu2025}. The presence of a cylindrical tank's vertical sidewall acts to stabilize the oscillatory bulk mode, thus raising the critical Rayleigh number $Ra_o^{cyl}$ for low $Pr$ oscillatory modes in a cylinder \citep{Liao_Zhao_Chang_2006, ZHANG_LIAO_2009}.

The other bulk RC mode is the steady ``geostrophic'' mode \citep{Niiler_Bisshopp_1965, JULIEN_KNOBLOCH_1998, Kunnen04052021}. It is stationary and emerges at higher Rayleigh numbers than the oscillatory modes in low $Pr$ RC \citep{Horn_Schmid_2017, aurnou2018rotating, Vogt_Horn_Aurnou_2021}.

 The transition from the conductive sidewall state to an asymmetric traveling-wave state occurs via a supercritical Hopf bifurcation for all $Pr$ fluids \citep[e.g.,][]{Zhong_Ecke_Steinberg1991, Goldstein_Knobloch_Mercader_Net_1994,Vasil2024}.
The critical $Ra$ and drift frequency of wall mode convection with no-slip boundary conditions are presented in \citet{Herrmann_Busse_1993,Liao_Zhao_Chang_2006}.

Rotating magnetoconvection also exhibits convective multimodality, including oscillatory, wall, magnetic and geostrophic modes \citep[e.g.,][]{sakuraba2002linear, aujogue2015onset, HornAurnou2022, HornAurnou2025}. Additionally, a distinct magnetostrophic mode emerges when the dominant force balance occurs between Lorentz and Coriolis forces. This coexistence of these multiple RMC modes occurs within a specific Elsasser number range, which was first identified by Donna Elbert and is referred to here as the Elbert range, following \citet{HornAurnou2022}.

The critical Rayleigh numbers and nondimensional frequencies associated with these modes are summarized in Table \ref{tab:conv_modes}. The frequencies are nondimensionalized by the system's rotation frequency, $f_{\Omega} = \Omega / (2 \pi)$ with $\Omega$ denoting the angular frequency in  \si{rad/s}.  These theoretical values serve as useful references for interpreting the onset behaviors and transitions between convection regimes observed in our experiments.


%
The oscillatory and wall mode onsets and frequencies for rotating magnetoconvection (RMC) are calculated in Table \ref{tab:conv_modes} using the Jupyter Notebook provided in \citet{HornAurnou2022}. 

\setlength{\tabcolsep}{4pt}      
\renewcommand{\arraystretch}{1.5}
\begin{table}[t!]
 \begin{center}
 \small
\def~{\hphantom{0}}
 \resizebox{\textwidth}{!}{
  \begin{tabular}{lll}
          Convection mode &  Critical Rayleigh number   & Mode frequency \\[3pt]
           \hline
     Oscillatory bulk mode & $Ra_o^{\infty}=17.4\left(Ek/Pr\right)^{-4/3}$    & $f_o^{\infty} =4.8\left(Ek/Pr\right)^{1/3}$ \\
       Steady geostrophic mode & $Ra_s^{\infty}=(8.68 - 9.63Ek^{1/6})Ek^{-4/3}$  & $-$ \\
        Wall mode & $Ra_w=31.8Ek^{-1}+46.6Ek^{-2/3}$  & $f_w = 131.8Ek/Pr-1464.5Ek^{4/3}/Pr$ \\
        Steady magnetostrophic mode & $Ra_{mac}^{\infty} = \left(\pi^2 (1 + \sqrt{1 + \Lambda^2})^2/\Lambda\right) Ek^{-1}$  & $-$ \\
  \end{tabular}
  }
  \caption{Critical Rayleigh number and nondimensional frequency (nondimensionalized by rotation rate) of oscllatory bulk mode, steady geostrophic mode and wall mode \citep{Horn_Schmid_2017} in rotating convection,  and steady magnetostrophic mode in rotating magnetoconvection \citep{HornAurnou2022}. The superscript $\infty$ denotes values in an infinite layer.}
  \label{tab:conv_modes}
  \end{center}
\end{table}

Figure \ref{fig:earthcore_linearprediction}b shows critical $Ra$ predictions for the different modes of RMC in Earth's outer core, made using core values of $Ek = 10^{-15}$ and $Pr = 0.1$. The semi-transparent blue shaded box describes the estimated ($Ra$-$\Lambda$) parameter space ($Ra\in[10^{23},10^{29}]$, $\Lambda\in[0.1,10]$) for RMC in Earth's outer core \citep[e.g.,][]{Gubbins2001, AurnouEPSL2003, Gillet2010, Roberts_2013, AbbateAurnou2023}. 
The magnetostrophic, oscillatory and geostrophic convection modes are all strongly supercritical within this region of parameter space; $Ra$ greatly exceeds each mode's estimated critical values everywhere in the ($\Lambda$, $Ra$) box. This supports the relevance of our multimodal laboratory experimental approach, as it likely mimics the essential dynamics of rotating magnetoconvection in Earth's outer core, which, based on Figure \ref{fig:earthcore_linearprediction}b, is also highly multimodal in nature. 

\subsection{Transport scalings}
Convective velocity scaling predictions can be derived by balancing the kinetic energy production and viscous dissipation terms in the mean kinetic energy equation \citep[e.g.,][]{ingersoll1982motion, Aubert_2001, Jones_2011, KingBuffet2013, Hawkins_2023}. A generalized expression for the Reynolds number ($Re$) can be obtained that describes the convective flow:
\begin{equation}
Re\sim\left(\frac{Ra(Nu-1)}{Pr^2}\right)^{1/2}\frac{\ell}{H},
\label{eq:u_cia}
\end{equation}
where $\ell$ is the characteristic convective flow scale \citep{KingBuffet2013}
and the Nusselt number, $Nu$, is the ratio of total and conductive heat transfer \citep[][]{spiegel1971convection, goluskin2016internally}.

In rotating convective flows, the geostrophic turbulence scale is employed:
\begin{equation}
\ell \sim Ro_c H \sim Ro^{1/2}H, 
\label{eq:l_cia}
\end{equation}
where $Ro = U/(2\Omega H)$ is the Rossby number \citep{CARDIN1994, Guervilly_2019, Aurnou2020}.  This choice of $\ell$ leads to the well-known CIA velocity scaling against which we will compare our RC and our RMC data in the following section: 
\begin{equation}
Re_{cia} = \left(\frac{Ra(Nu-1)}{Pr^2}\right)^{2/5}Ek^{1/5}.
\label{eq:Re_cia}
\end{equation}

The CIA scaling is inviscid by construction.  However, it may not be free of diffusive effects, a.k.a., fully diffusivity-free (DF), as the heat transport may still depend on the thermal diffusive properties of the fluid. To address this, we use \citet{JulienKeith2012}'s diffusivity-free scaling for rotating convective heat transport,
\begin{equation}
Nu-1=C_J \, Ra^{3/2} \, Ek^2 \, Pr^{-1/2}= C_J \, Ro_c^2 \, Pe_{ff},  
\label{eq:DF_Nu}
\end{equation}
where $C_J\approx1/25$ is a constant derived from their asymptotically-reduced simulations of rapidly rotating convection. Here, $Pe_{ff}= U_{ff} H / \kappa = \sqrt{Ra\, Pr}$ is the free-fall Péclet number, which estimates the ratio of non-rotating thermal advection and thermal diffusion \citep{Aurnou2020}. The corresponding DF velocity scaling follows by substituting (\ref{eq:DF_Nu}) into (\ref{eq:Re_cia}), yielding \citep{abbate2024, Hadjerci_Bouillaut_Miquel_Gallet_2024, abbate2024diffusivityfree}:
\begin{equation}
Re_{df} = C_J^{2/5}Ro_c^{2}Ek^{-1}= C_J^{2/5}Ro_cRe_{ff} \, , 
\label{eq:ReDF}
\end{equation}
where $C_J^{2/5} \approx 0.276$.

A separate heuristic velocity scaling was postulated in \citet{KingBuffet2013} to describe magnetostrophic convection.  However, we have been unable to collapse any of our velocity data with this $Re_{mac}$ scaling, 
and, thus, will not consider it further in this study.


\section{Methods}
\label{method}

\subsection{Laboratory set-up and diagnostics}
\label{subsec:set-up}

\begin{figure}[t!]
  \centerline{\includegraphics[width=14cm]{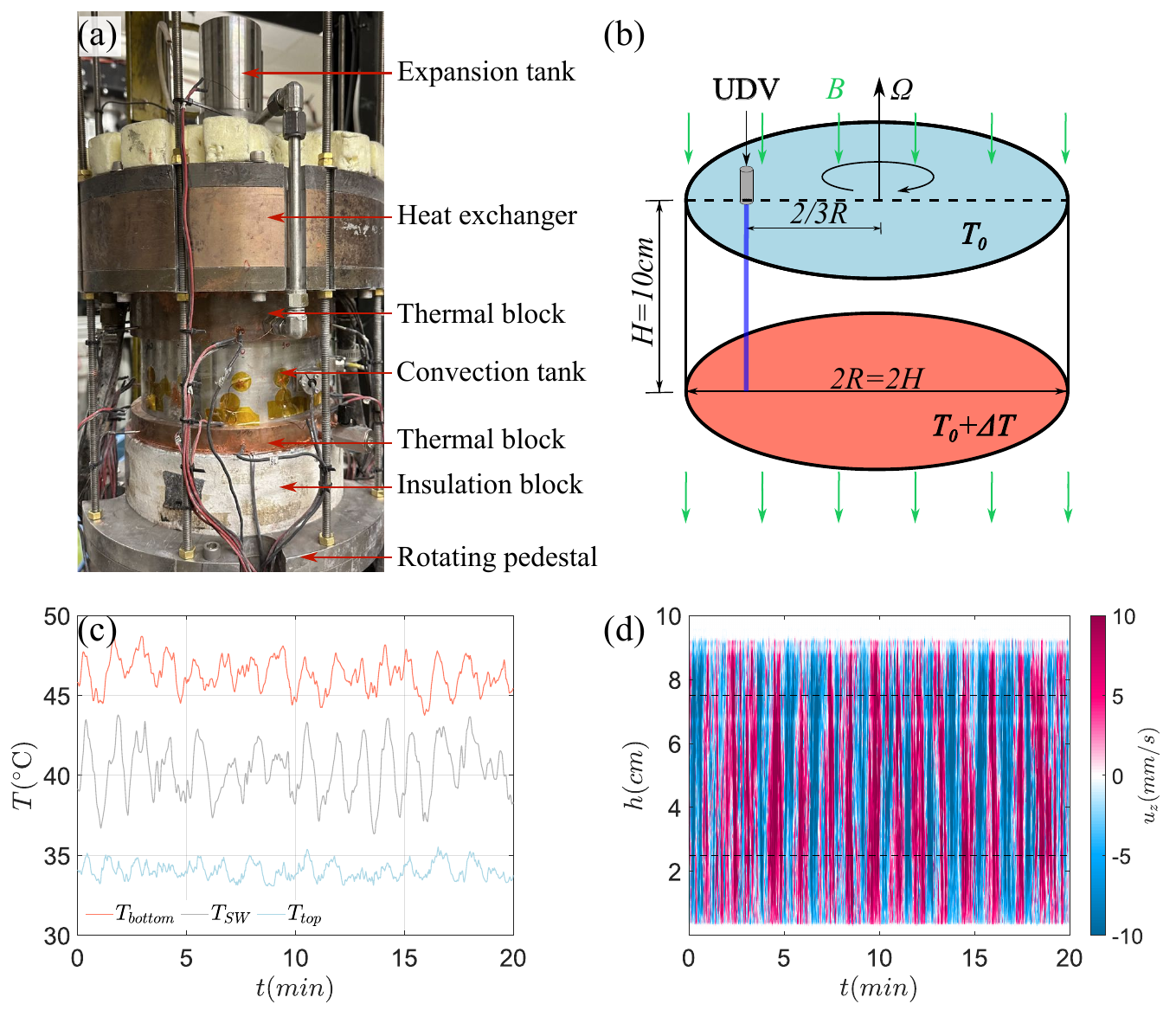}}
  \caption{(a) A photograph of the RoMag set-up with aspect ratio $\Gamma = 2R/H=2$ tank ($H=9.84$ cm; $R=9.84$ cm) used here. (b) schematic of the convection cell showing the ultrasonic Doppler velocimetry (UDV) measurements and the imposed temperature, rotation and magnetic fields; (c,d) examples of temperature and velocity signals obtained from an RC case carried out at $Ra = 3.08\times 10^6$ ($P=780$ W) and $Ek=10^{-4}$ ($1.78$ rpm).}
\label{fig:photosetup_examplemeasurements}
\end{figure}

Figure~\ref{fig:photosetup_examplemeasurements}a shows a photograph of the experimental set-up, `RoMag'. The test section consists of top and bottom copper lids and a cylindrical stainless steal sidewall section situated between the lids. The bottom lid is heated by a basal heat pad, which provides an input heating power of $P=0$ to $780$ W. The applied heating is removed from the top lid by a heat exchanger connected to a recirculating thermostated chiller. This equipment is mounted on a rotating pedestal. The test section is insulated with a 1 cm thick layer of aerogel, which has a thermal conductivity of approximately $k = 0.015$ W/(m K). The working fluid is gallium, which has a Prandtl number $Pr	\approx 0.026$.  Gallium's thermal conductivity is $k = 31.4$ W/(m k) and the kinematic viscosity of the gallium is $\nu = 3.4 \times 10^{-7}$ m$^2$ / s \citep{aurnou2018rotating}.  For further RoMag device details, see \citet{king2012heat}. 

Figure \ref{fig:photosetup_examplemeasurements}b contains a schematic of the test section. The dimensions of the fluid volume are $H=9.84$ cm in height and $R=9.84$ cm in radius, yielding $\Gamma = 2R/H =2$. A nearly uniform vertical magnetic field is applied by an external solenoid, generating a magnetic flux density of up to $0.2$ T. 
The tank is spun here between 0 and 36 revolutions per minute (rpm) in a left-handed sense. Due to this left-handed applied rotation, a left-handed ($s$, $\phi$, $z$) coordinate system is used here, with $z = 0$ cm defining the base of the fluid layer. The top and bottom lids each contain six embedded thermistors used to measure the bottom to top temperature difference $\Delta T = (T_{bottom} - T_{top})$ across the fluid layer. With this information, the Nusselt number can be computed as:
\begin{equation}
Nu=qH/(k \Delta T) , 
\label{eq:Nu}
\end{equation}
where $q = P/(\pi R^2)$ is the heat flux from the basal heat pad, corrected for any sidewall losses. The midplane ($z = H/2$) temperature is monitored by 12 equally spaced thermistors situated on the exterior face of the sidewall. 
Examples of the individual thermistor signals obtained from the bottom lid ($T_{bottom}$), the side-wall ($T_{SW}$) and the top lid ($T_{top}$) are shown in Figure~\ref{fig:photosetup_examplemeasurements}c.

Convective flow velocities in the liquid metal are experimentally measured using the non-invasive ultrasonic Doppler velocimetry (UDV) technique \citep{Takeda_1995, brito2001ultrasonic}. Yielding a chord of velocity data along the UDV transducer's line of sight (Figure \ref{fig:photosetup_examplemeasurements}b), UDV has been successfully applied to obtain velocity profiles in fluids that are both transparent \citep[e.g.,][]{noir2001experimental, brito2004turbulent} and opaque \citep[e.g.,][]{Aubert_2001, Vogt2018, Zürner_2019, Vogt_Horn_Aurnou_2021, Cheng_Mohammad_Wang_Keogh_Forer_Kelley_2022}. In this study, we employ the DOP5000 UDV system (Signal Processing SA, Switzerland), operating at an emission frequency of $f_e = 8$~\si{MHz}, and equipped with a vertically-oriented piezoelectric ceramic transducer. This UDV probe is positioned in the top lid at $2R/3$ from the container's center to measure the instantaneous vertical velocity component $u_z$ across the entire fluid layer height. A Hovm\"{o}ller diagram of UDV $u_z$ data is plotted in Figure \ref{fig:photosetup_examplemeasurements}d.

The root mean square (RMS) value of the vertical UDV velocity, $u_{z,rms}$, is evaluated over time and over the height range from 0.25H to 0.75H (see the two horizontal dashed lines in Figure~\ref{fig:photosetup_examplemeasurements}d):
\begin{equation}
u_{z,\mathrm{rms}} = \sqrt{\frac{1}{N_t N_z} \sum_{i=1}^{N_t} \sum_{j=1}^{N_z} u_z^2(z_j, t_i)}.
\label{eq:uzrms}
\end{equation}
\noindent This then allows us to calculate the experimentally measured axial Reynolds number as:
\begin{equation}
Re_z = \frac{u_{z,\mathrm{rms}}H}{\nu}.
\label{eq:Rez}
\end{equation}
Following that, the experimentally measured Rossby number is calculated via:
\begin{equation}
Ro_z =\frac{u_{z,\mathrm{rms}}}{2\Omega H}=\frac{u_{z,\mathrm{rms}}H}{\nu}\frac{\nu}{2\Omega H^2} = Re_zEk.
\label{eq:Rez}
\end{equation}
We estimate the peak axial velocity, $u_{z,\max}$, by identifying the maximum value of the height-averaged velocity:
\begin{equation}
u_{z,\max} = \max\left(\frac{1}{N_z}\sum_{j=1}^{N_z} u_{z}(z_j, t_i)\right), 
\label{eq:uzmax}
\end{equation}
which we will use in (\ref{eq:N_ell}) to calculate $N_\ell$. Since the axial velocities do not greatly vary in value in $z$ this scheme acts mainly to filter out any spurious high velocity values while retaining physically-meaningful peak velocity signatures.

\begin{figure}[htbp!]
  \centerline{\includegraphics[width=13.5cm]{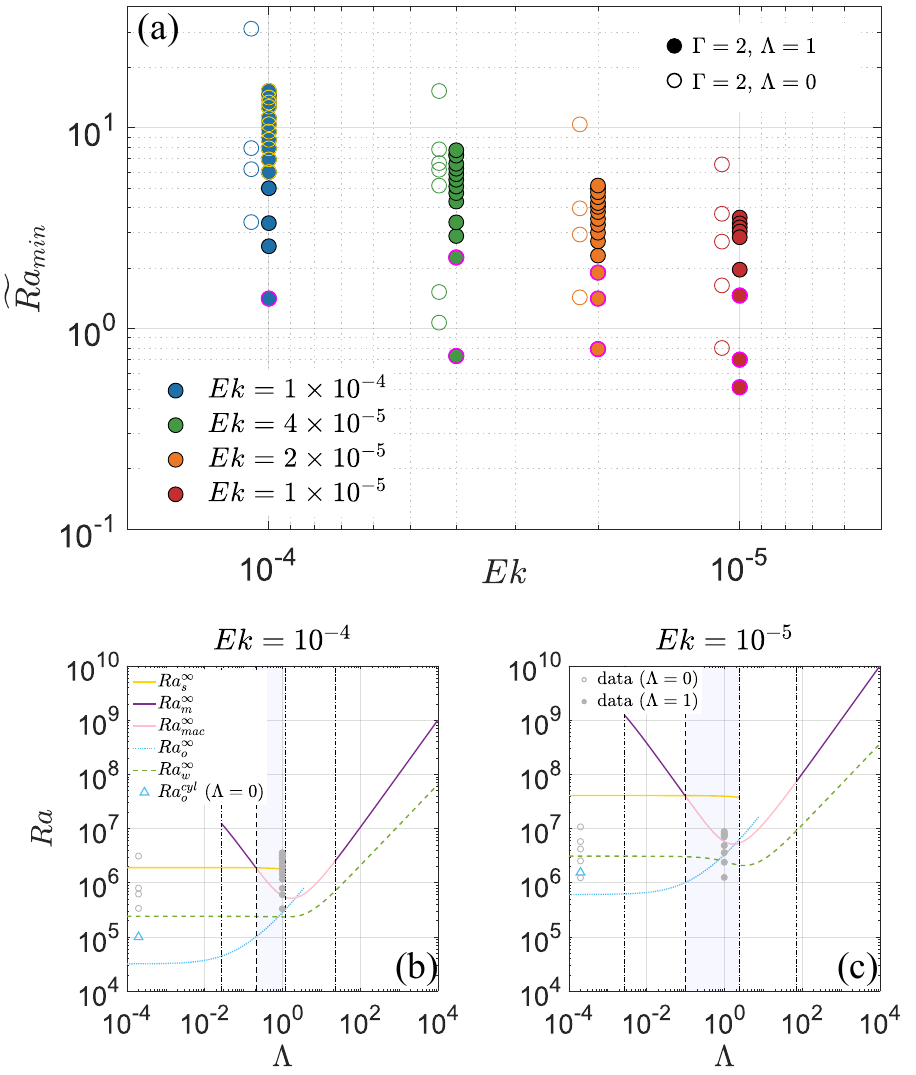}}
  \caption{(a) Experimental ($Ek$, $Ra/Ra_{crit,min}$) parameter space, where $Ra_{min}$ is the minimum of all the critical $Ra$ values. Hollow RC symbols are offset in $Ek$ for visibility.  Gold edge color indicates cases in geostrophic regime, magenta edge color marks cases where the magnetostrophic mode is not active.
  Liquid gallium critical $Ra$ predictions from \citet{HornAurnou2022} for (b) $Ek = 10^{-4}$ and (c) $Ek = 10^{-5}$. Hollow (filled) circles denote RC (RMC) experiments. 
  }
\label{fig:parameter_space}
\end{figure}

\subsection{Parameter space}
\label{subsec:parameter_space}
\sloppy
Figure~\ref{fig:parameter_space}a plots the experimental parameter space investigated.  Hollow circles correspond to $\Lambda = 0$ RC cases, whereas solid circles denote $\Lambda = 1$ RMC cases. Symbol color denotes the experimental rotation rates, corresponding to Ekman numbers of $Ek= 10^{-4}$,  $4\times 10^{-5}$, $2\times 10^{-5}$ and $10^{-5}$. The RC cases have all been slightly displaced to the left so that they are not hidden in the plot by the solid RMC symbols. 

At each $Ek$ value, the input power $P$ was systematically changed in order to vary the buoyancy forcing ($Ra$). The convective supercriticality of each case, $\widetilde{Ra}_{min}$, is shown on the y-axis.  Here we define supercriticality as 
\begin{displaymath}
    \widetilde{Ra}_{min} = Ra/Ra_{crit,min}
\end{displaymath}
where
\begin{displaymath}
    Ra_{crit,min} = min\{Ra_o^{\infty},Ra_o^{cyl},Ra_w^{\infty}, Ra_s^{\infty}, Ra_{mac}^{\infty} \}.
\end{displaymath}
In the RC cases, $Ra_{mac}^{\infty}$ is not considered in calculating $\widetilde{Ra}_{min}$.


Figures~\ref{fig:parameter_space}b and \ref{fig:parameter_space}c show the different modes critical $Ra$ curves as a function of the Elsasser number $\Lambda$ for liquid Gallium ($Pr=0.025$) at $Ek=10^{-4}$ and $Ek=10^{-5}$, respectively, all computed via \cite{HornAurnou2022}'s Jupyter notebook. Additionally, the cyan triangles mark $Ra_o^{cyl}$ in a $\Gamma=2$ cylindrical tank \citep{ZHANG_LIAO_2009}. In both these panels, with increasing $Ra$, RC first develops via the bulk oscillatory mode (cyan dotted line / cyan triangle), followed by the development of the wall mode (green dashed line), and finally the stationary geostrophic mode (gold solid line). In contrast, $\Lambda=1$ RMC has a different sequence of modal development. RMC initially manifests as a wall mode, followed nearly immediately by the emergence of the oscillatory mode, then the stationary magnetostrophic mode (pink solid line), and finally the stationary geostrophic mode. 

Input and output parameter values for all the experimental cases are provided in the data tables in the Appendix.

\begin{figure}[b!]
  \centerline{\includegraphics[width=15cm]{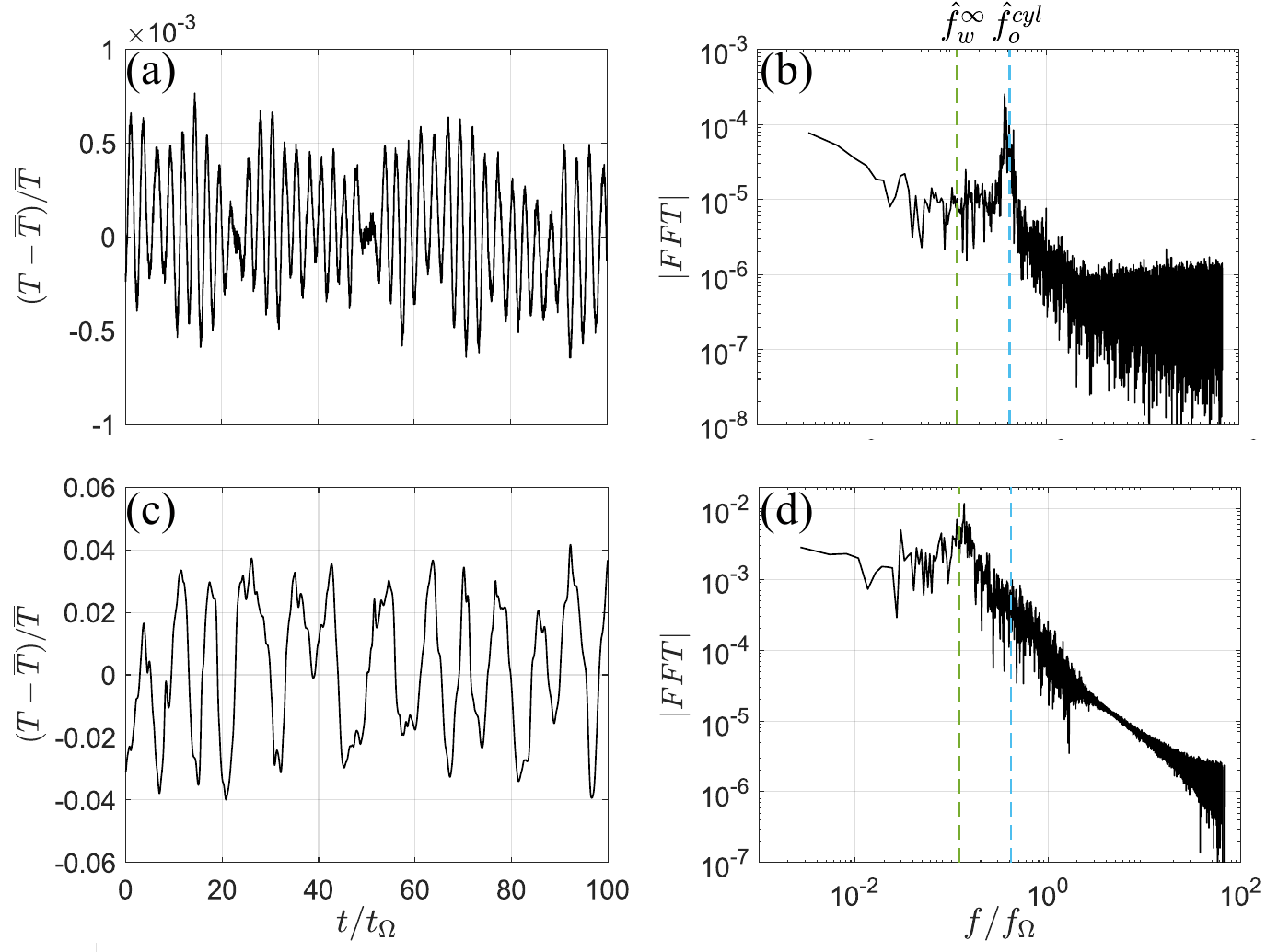}}
  \caption{Validation cases for side-wall thermistor measurements of oscillatory and wall modes  at $Ek=4\times 10^{-5}$ and $\Lambda = 0$: ($a$,$c$) normalized side wall temperature (located at at $270^\circ$) fluctuations at $Ra=4.41\times 10^{5}$ ($Ra/Ra_{o}^{\infty}=4.4$, $Ra/Ra_{w}=0.64$) and $Ra=1.93\times 10^{6}$ ($Ra/Ra_{o}^{\infty}=19.3$, $Ra/Ra_{w}=2.8$); ($b$,$d$) corresponding amplitude of the Fourier Transform of temperature fluctuations in ($a$,$c$).} 
\label{fig:validation_case}
\end{figure}

\subsection{Validation cases}
\label{Validationcase}
Figure~\ref{fig:validation_case} presents sidewall thermal measurements for two $Ek= 4\times10^{-5}$ RC cases conducted to assess the accuracy and reliability of our experimental configuration under different convective regimes.  In the first case, $Ra = 4.4 \times 10^5$, which just exceeds the onset value for the oscillatory mode in a cylindrical geometry ($Ra/Ra_{o}^{\infty}=4.4$, $Ra/Ra_{w}=0.64$). In the second case, $Ra = 1.9 \times 10^6$ such that it wall modes are active as well ($Ra/Ra_{o}^{\infty}=19.3$, $Ra/Ra_{w}=2.8$).

Figures \ref{fig:validation_case}a and \ref{fig:validation_case}c shows the time series of normalized temperature fluctuations, $\left(T-\bar{T}\right)/\bar{T}$, measured by the mid-height side-wall thermistor positioned at $270^o$, where $\bar{T}$ is the mean value of the time series. The time axes are nondimensionalized by the system's rotation period $t_{\Omega} = 2 \pi / \Omega$. Both signals exhibit periodic oscillations. However, the bulk oscillations in (a) are higher frequency and are have a very low amplitude when measured on the exterior sidewall.  In contrast, the wall modes are lower frequency and are far larger in amplitude when measured on by these exterior sidewall situated sensors. 

Corresponding Fourier spectra are shown in Figures \ref{fig:validation_case}b and \ref{fig:validation_case}d with the frequency axis nondimensionalized using the system's rotation frequency $f_{\Omega} = t_\Omega^{-1} = \Omega / (2 \pi)$. In (b), a strong spectral peak aligns with the predicted oscillation frequency for cylindrical geometry, $\hat{f}_o^{cyl}=f_o^{cyl}/f_{\Omega}$ (vertical cyan dashed line). In (d), the wall mode generates the dominant spectral peak near the predicted wall mode frequency for a semi-infinite plane layer, $\hat{f}_w^{\infty}=f_w^{\infty}/f_{\Omega}$ (vertical green dashed line). These cases successfully validate that our laboratory set-up generates convective flows that are in good agreement with the predictions for low $Pr$ RC, and, further, that the characteristics of different convective modes are detectable and distinguishable.


\section{Results}
\label{Results}

\subsection{Flow fields}
\label{Comparison}

\begin{figure}[t!]
\centerline{\includegraphics[width=15cm]{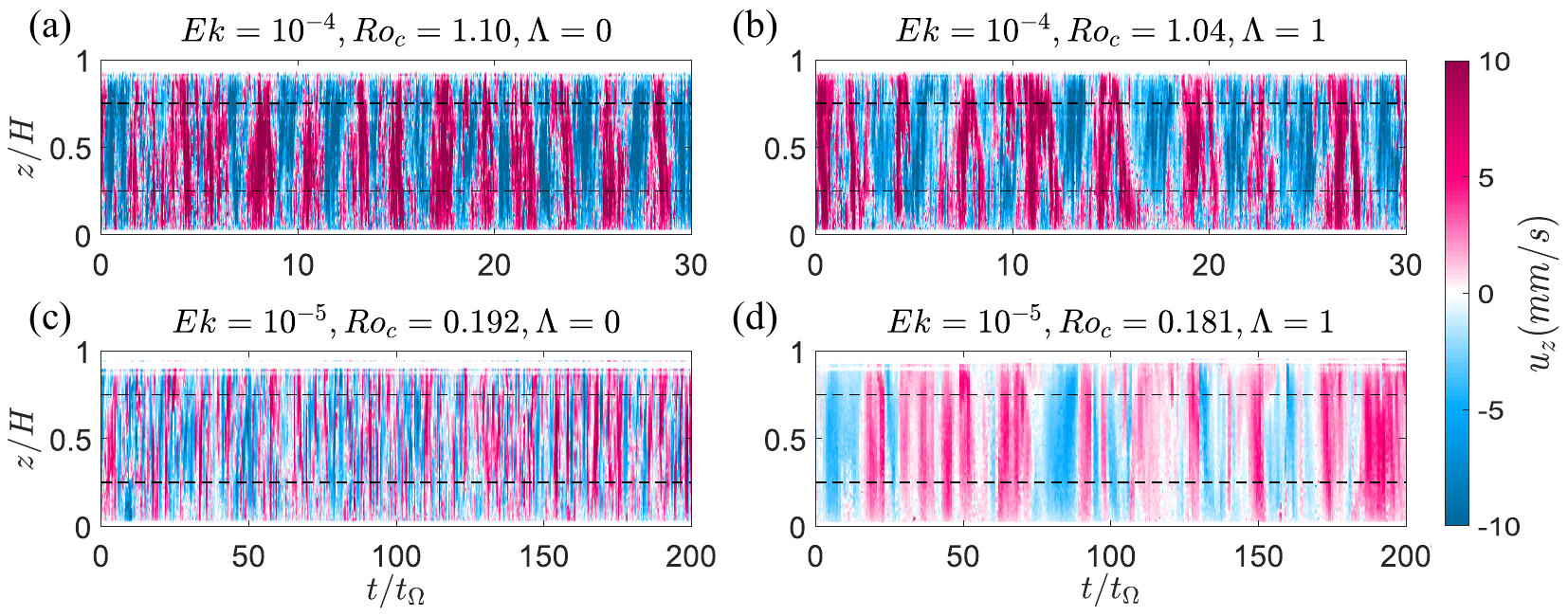}}
  \caption{Hovm\"{o}ller diagrams of vertical velocities obtained from the vertical UDV probe. Left column $\Lambda=0$ and right column $\Lambda=1$; top row $Ek=10^{-4}$ and bottom row $Ek=10^{-5}$; input power $P=780W$ for all cases; Rayleigh numbers: 
  (a) $Ra=3.08\times10^{6}$; (b) $Ra=3.57\times10^{6}$ ; (c) $Ra=1.01\times10^{7}$ ; (d) $Ra=8.80\times10^{6}$.}
  \label{fig:velocity_diagram}
\end{figure}

Figure \ref{fig:velocity_diagram} shows the spatiotemporal evolution of vertical velocity $u_z$ measured using the vertical 
UDV for four different cases. The left column corresponds to rotating convection (RC) at Ekman numbers $Ek=10^{-4}$ and $Ek=10^{-5}$, while the right column represents RMC under the same Ekman numbers but with $\Lambda = 1$. The vertical axis denotes the spatial coordinate along the tank height (H), while the horizontal axis represents the dimensionless rotational time ($t/t_{\Omega}$). The details of the supercriticality and parameter values for these cases are provided in Table \ref{tab:fig5_6}. 

A comparison of Figures \ref{fig:velocity_diagram}a and \ref{fig:velocity_diagram}b reveal that both cases exhibit flow oscillations of comparable amplitude, spanning the system scale (H). However, the presence of the magnetic field in the RMC yields a lower oscillation frequency. 
Both cases are unstable to the geostrophic mode, with supercrticality $\widetilde{Ra}_{s}^{\infty}=1.6$ and $2.0$, respectively, as predicted by linear theory for an infinite plane layer. However, the flow remains predominantly governed by the oscillatory mode. This discrepancy may arise from an underestimation inherent in the linear theoretical planar prediction, relative to that of an actual cylinder. 
Although the convective Rossby number for both $Ek = 10^{-4}$ cases is approximately unity ($\sim 1$), their spectra each show a peak consistent with oscillatory behavior. The $\Lambda=0$ case exhibits a strong peak near the predicted oscillatory RC frequency, while the $\Lambda=1$ case also shows an oscillatory peak, albeit weaker and shifted to a lower frequency, consistent with a slowed RMC oscillation prediction (see Figure~\ref{fig:FFT_velocity_SWtemperature_comparison} in the Appendix).


\setlength{\tabcolsep}{4pt}      
\renewcommand{\arraystretch}{0.9} 

\begin{table}[t!]
  \centering
  \small
  \resizebox{\textwidth}{!}{%
  \def~{\hphantom{0}}
  \begin{tabular}{ccccccccccccc}
    Fig.~5 & $\Lambda$ & $Ek\cdot10^{4}$ & $Ra\cdot10^{-6}$ & $Ro_c$ 
             & $\widetilde{Ra}_o^{\infty}$ & $\widetilde{Ra}_o^{cyl}$ 
             & $\widetilde{Ra}_w^{\infty}$ & $\widetilde{Ra}_{mac}^{\infty}$ 
             & $\widetilde{Ra}_{s}^{\infty}$ & $Nu$ & $Re_{z}$ & $Ro_{z}$ \\[2pt]
    \hline
    (a) & 0 & 1   & 3.08  & 1.04  & 97.0 & 31.2 & 12.9 & $-$ & 1.6 & 6.69 & 2565 & 0.2565 \\
    (b) & 1 & 1   & 3.57  & 1.10  & 12.7 & $-$  & 15.3 & 6.2 & 2.0 & 5.96 & 2005 & 0.2005 \\
    (c) & 0 & 0.1 & 10.10 & 0.192 & 16.7 & 6.6  & 3.3  & $-$ & 0.2 & 2.14 & 1289 & 0.0129 \\
    (d) & 1 & 0.1 & 8.60  & 0.181 & 2.7  & $-$  & 3.6  & 1.5 & 0.2 & 2.47 &  691 & 0.0069 \\
  \end{tabular}
  } 
  \caption{Input and output parameters for the cases shown in Figure~\ref{fig:velocity_diagram}. 
  The definitions of these parameters are provided in Sections~\ref{sec:nd_parameters}, \ref{subsec:set-up} and \ref{subsec:parameter_space}.}
  \label{tab:fig5_6}
\end{table}


Figure panels \ref{fig:velocity_diagram}c and \ref{fig:velocity_diagram}d allow us to compare the RC and RMC Hovm\"{o}ller UDV diagrams at a lower Ekman number, $Ek=10^{-5}$, over a 200 rotation period time window. Comparing the two cases, it is evident that the imposition of a vertical magnetic field, such that $\Lambda=1$, lowers the intensity of the RMC flow and causes it to become better aligned in the axial direction. Thus, a larger-scale flow structure emerges in the RMC case, consistent with findings from numerical simulations of Rayleigh-Bénard convection with a vertically imposed magnetic field \citep[e.g.,][]{Yan_2019}. 
There is no fundamental change in the gross flow structures in all four Hovm\"ollers: the axial velocities generally remain locally well aligned with the axis of the rotating tank $\hat{\mathbf{z}}$. These results are quantitatively consistent with, and compare well to, the laboratory-numerical liquid metal RC experiments of \cite{Vogt_Horn_Aurnou_2021}. It is interesting that this axiality appears to remain intact, although the convective Rossby numbers are near unity in the higher $Ek$ cases, while being closer to 0.2 in the lower $Ek$ cases. However, for the low $Ro_c$ $\Lambda = 1$ case, it is clear that slower, larger-scale, coherent motions dominate the system, in qualitative agreement with magnetostrophic theory and with simulation results such as those of \citet{Yadav_2016_PNAS}.





\subsection{Heat and momentum transport}
\label{global_transport}


\subsubsection{Scaling with supercriticality}
\label{supercriticalityscaling}

\begin{figure}[t!]
\centerline{\includegraphics[width=15cm]{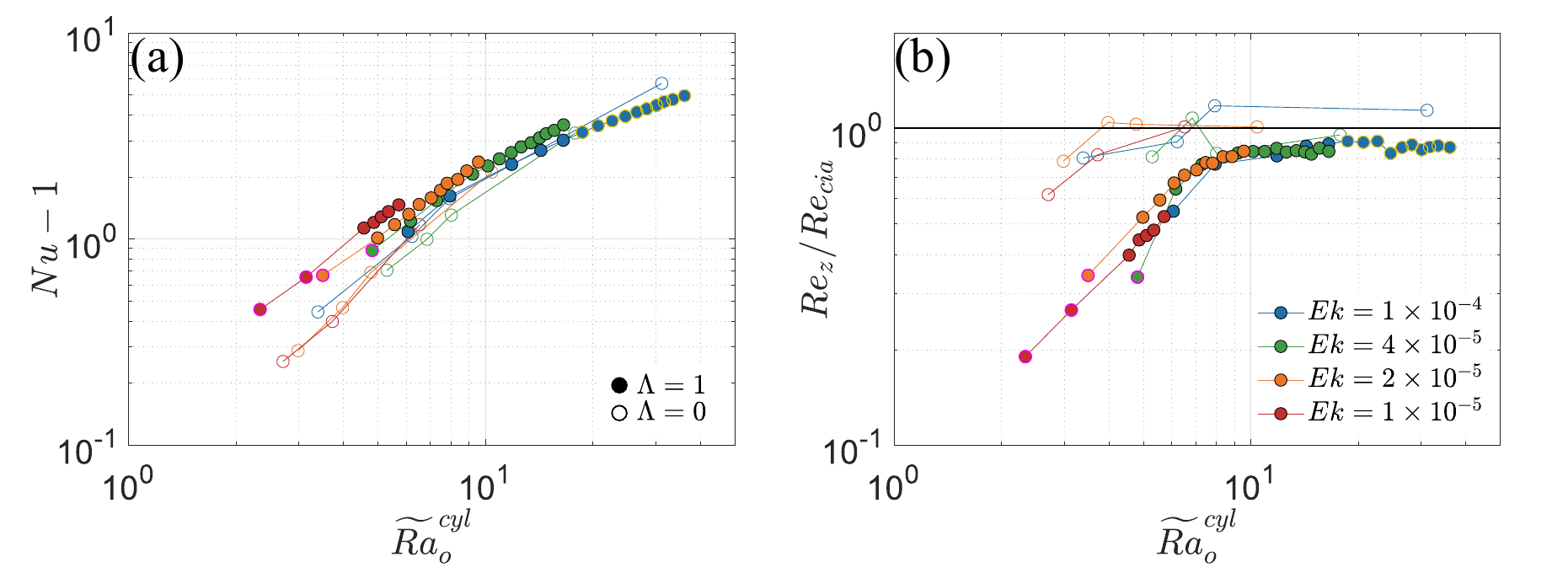}}
  \caption{(a) Convective heat flux relative to conduction heat flux $Nu-1$ 
  and (b) vertical Reynolds number normalized by the CIA scaling prediction, $Re_z/Re_{cia}$, as a function of the cylindrical supercriticality, $\widetilde{Ra}_{o}^{cyl}$ for rotating convection data (hollow circle) and rotating magnetoconvection data (filled circle). Gold edge color indicates cases in geostrophic regime, magenta edge color marks cases where the magnetostrophic mode is not active.}
\label{fig:Supercriticality}
\end{figure}

Figure~\ref{fig:Supercriticality}a shows the convective heat transfer, $Nu - 1$, as a function of supercriticality, $\widetilde{Ra} = Ra/Ra_o^{cyl}$, for rotating convection (hollow circles) and rotating magnetoconvection (filled circles). All the heat transfer data is moderately well collapsed as a function of $\widetilde{Ra}_o^{cyl}$. For $Ek < 10^{-4}$, the RMC $\Lambda = 1$ heat transfer is enhanced relative to the RC cases, and this enhancement becomes more pronounced as the Ekman number $Ek$ decreases. These observations are consistent with previous experimental \citep{KingAurnou2015PNAS,Grannan_2022} and numerical \citep{yadav2016effect} results. It was argued in \citet{Grannan_2022} that this enhancement was due to the presence of magnetostrophic wall modes \citep{Grannan_2022}. Based on Figure \ref{fig:velocity_diagram}d, it appears plausible that the $\Lambda = 1$ heat transfer enhancement here is due to large-scale, slightly lower velocity, coherent bulk flows \citep[cf.][]{chong2017, xia2023tuning}.

Figure~\ref{fig:Supercriticality}b presents the vertical Reynolds number $Re_z$ normalized by the CIA-predicted Reynolds number $Re_{cia}$, where the characteristic flow scale is taken to be the experimentally-measurable $\ell \sim Ro^{1/2}H$ \citep{CARDIN1994, Guervilly_2019}. For $\widetilde{Ra}_o^{cyl} \gtrsim 8$, the RC and RMC data collapse well and tend toward $Re_{cia}$ (solid black horizontal line). In contrast, we find for $\widetilde{Ra}_o^{cyl} < 8$ that the RC data remain close to $Re_{cia}$, whilst the RMC data fall significantly below $Re_{cia}$. We argue that this is due to magnetic damping. Overall, the thermovelocimetric data in Figures \ref{fig:velocity_diagram} and \ref{fig:Supercriticality} support the fundamental predictions for magnetostrophic convection: coherent, larger-scale, magnetically-damped flows act to enhance the global heat transfer.

As the magnetostrophic mode is essentially inertia-less, inertial forces are expected to be small compared to the Coriolis and Lorentz forces \citep{Roberts_2013,HornAurnou2022}. To quantify the relative importance of inertia, we turn to the local interaction parameter $N_{\ell}$ and  the convective interaction parameter $N_{c}$, respectively defined in equations \eqref{eq:N_ell} and \eqref{eq:N_c}. To compute $N_{\ell}$, we use the maximum vertical velocity $u_{z,max}$ defined in equation \eqref{eq:uzmax}. 
 This choice of velocity is better suited for capturing the dynamics of oscillatory flows. 
 
\begin{figure}[t!]
\centerline{\includegraphics[width=15cm]{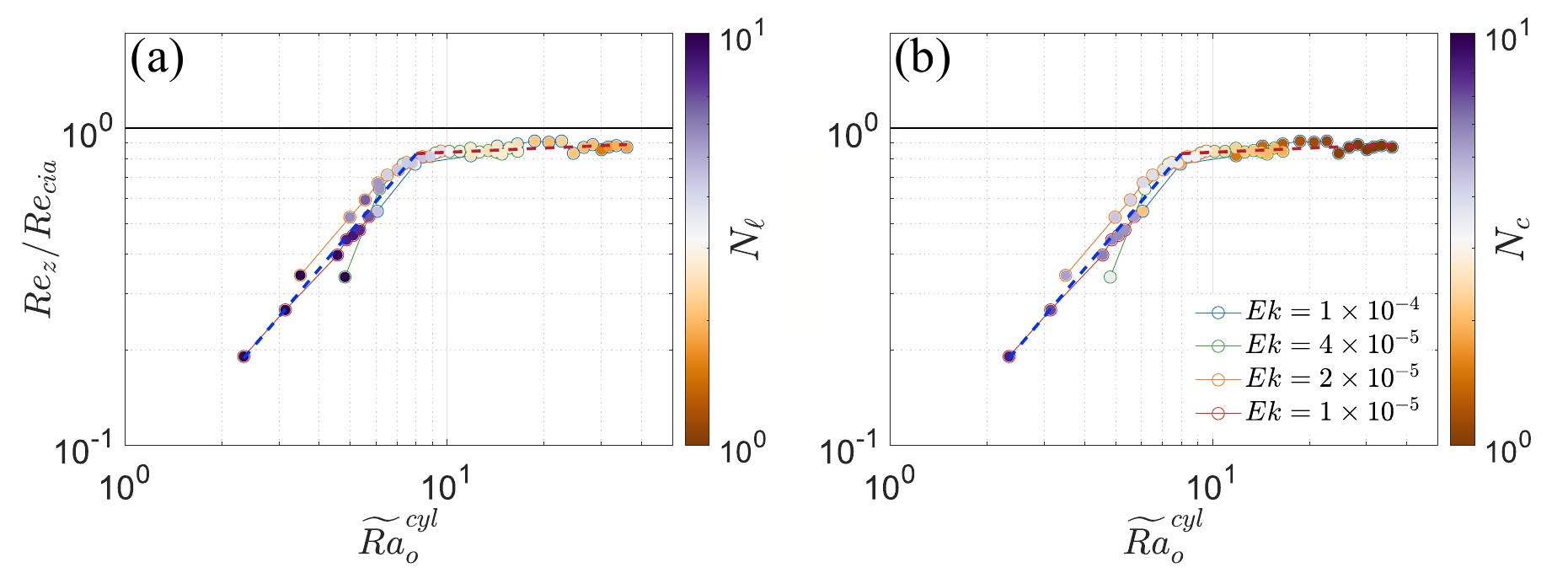}}
  \caption{Vertical Reynolds number normalized by the CIA scaling prediction, $Re_z/Re_{cia}$, plotted as a function of the cylindrical superficiality parameter, $\widetilde{Ra}_{o}^{cyl}$. The data are the same rotating magnetoconvection (RMC) measurements shown in Figure~\ref{fig:Supercriticality}, with marker edge colors indicating the Ekman number $Ek$, and marker fill colors representing: (a) the local interaction paramete $N_{\ell}$, and (b) the convective interaction parameter $N_{c}$. The blue dashed line shows a power-law fit to data with $\widetilde{Ra}_{o}^{cyl}<8$, while the dark red dashed line corresponds to a fit over the range $\widetilde{Ra}_{o}^{cyl}\geq8$. The intersection of the two fits give the transition values of $N_{\ell}\approx4.0$ and $N_c\approx3.1$.} 
\label{fig:Supercriticality_N}
\end{figure}

Figure \ref{fig:Supercriticality_N} shows  $Re_z/Re_{cia}$ plotted as a function of supercriticality $\widetilde{Ra}_o^{cyl}$, similar to Figure \ref{fig:Supercriticality}b.
However, here the symbol fill color represents either (a) $N_{\ell}$ or (b) $N_{c}$, and the edge color indicates the Ekman number $Ek$. We observe that both $N_{\ell}$ and $N_{c}$ approach values near 3 at the knick point near $\widetilde{Ra}_o^{cyl} \approx 8$. To estimate the critical $N$ values, we perform separate power-law fits to the data for $\widetilde{Ra}_o^{cyl}<8$ and $\widetilde{Ra}_o^{cyl}\geq8$ (shown as blue and red dashed lines in Figure \ref{fig:Supercriticality_N}). The averaged values near the intersection point of the two fits give transitions values of $N_{\ell} \approx 4.0$ and $N_{c} \approx 3.1$, indicating that both parameters are on the order of unity. The critical values of the interaction parameters indicate a transition from a magnetostrophic regime above the transition value to geostrophic turbulence below the transition value. 
Both the local interaction parameter $N_{\ell}$ and the convective interaction parameter $N_c$ exhibit a similar critical threshold value of around 3 to 4. While $N_{\ell}$ is defined based on the measured convective velocity (an output quantity), $N_c$ is formulated using the input control parameters.
%

 \begin{figure}[t!]
\centerline{\includegraphics[width=12cm]{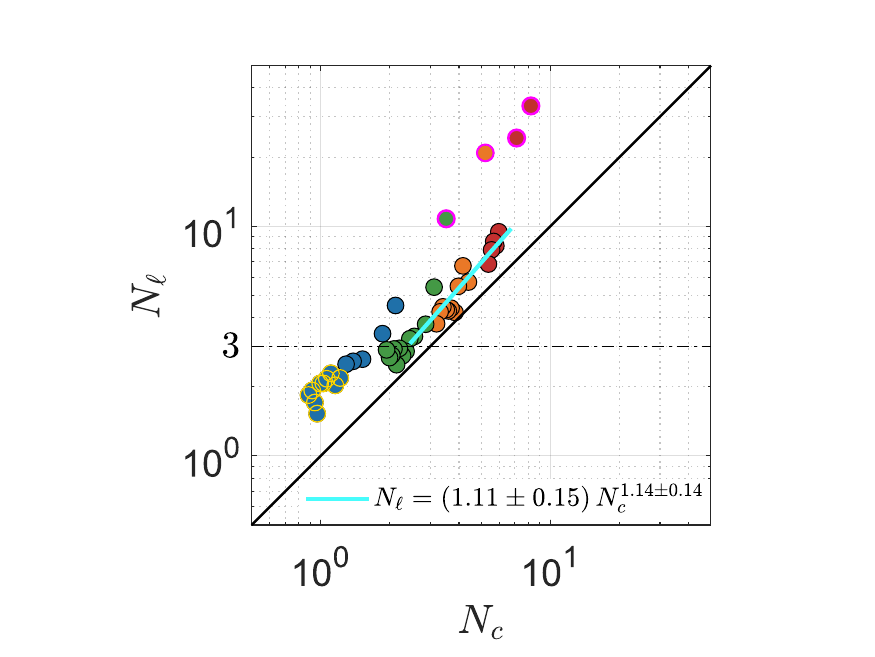}}
  \caption{Local interaction parameter $N_{\ell}$ vesus convective interaction parameter $N_{c}$ for RMC data, gold edge color indicates cases in geostrophic regime, magenta edge color marks cases where the magnetostrophic mode is not active, cyan dashed line is the best of the data with $N_{\ell}\ge3$ and magnetostrophic unstable (magenta edge color data excluded), black solid line represents $N_{\ell}=N_c$, black dot-dashed line indicates $N_{\ell}=3$.}
\label{fig:N_ell_vs_N_c}
\end{figure}

 Figure \ref{fig:N_ell_vs_N_c} presents $N_{\ell}$ as a function of $N_c$, enabling direct comparison between these two interaction parameter definitions. The figure reveals that $N_{\ell}$ closely follows $N_c$ across most RMC cases, with the exception of data points that are near the onset point of magnetostrophic convection, which are highlighted with magenta edge colors in Figure \ref{fig:N_ell_vs_N_c}. To quantify their relationship, we performed a power-law fit to the data with $N_{\ell}\ge3$, yielding the empirical relation $N_{\ell}=\left(1.11\pm 0.15\right)N_c^{1.14\pm0.14}$. This strong, approximately linear correspondence between $N_{\ell}$ and $N_c$ across most cases suggests that the convective interaction parameter, which is based solely on input parameters, can serve as a reliable proxy for characterizing the flow regime relatively near $N_\ell \approx N_c \gtrsim 3$. However, since $N_{\ell}$ can be directly obtained from our measurements and makes no assumptions on flow velocities or scales, we will use it to predict the Reynolds number under Earth's core conditions in Section \ref{sec:Application}.

\subsubsection{Diffusivity-free scaling}

\begin{figure}[t!]
\centerline{\includegraphics[width=13.5cm]{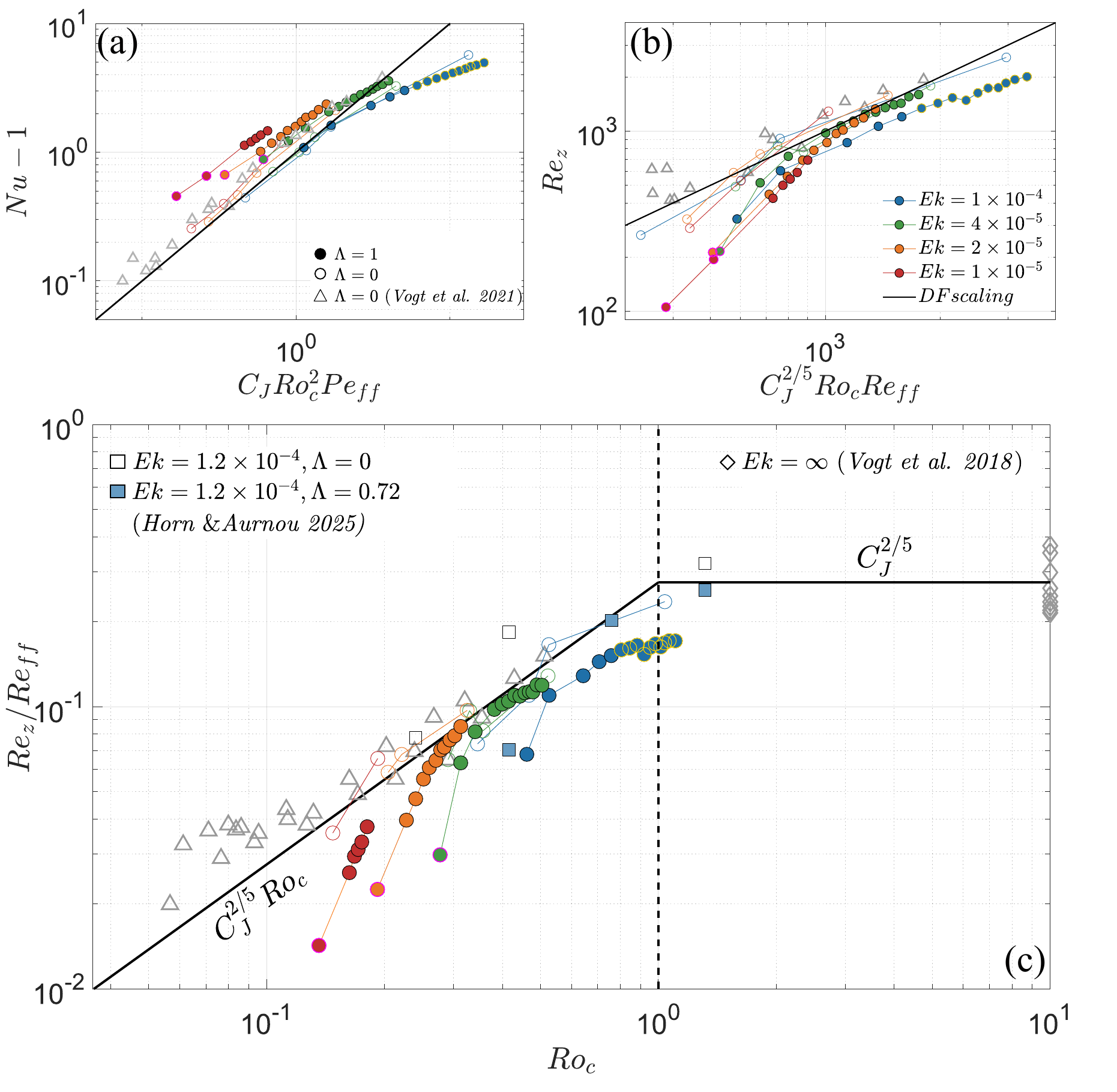}}
  \caption{($a$) Convective heat flux relative to conduction heat flux $Nu-1$ as a function of the diffusivity free prediction $C_JRo_c^2Pe_{ff}$; ($b$) vertical Reynolds number $Re_z$ as a function of diffusivity free prediction $C_J^{2/5}Ro_cRe_{ff}$; ($c$) vertical Reynolds number $Re_z$, normalized by the free-fall Reynolds number $Re_{ff}$, plotted as a function of the convective Rossby number $Ro_c$. The diagonal black solid line shows the diffusivity-free prediction for $Ro_c \ll 1$ and the horizonal black solid line is the diffusivity-free prediction for $Ro_c \gg 1$ \citep{Aurnou2020}. Filled circles with pink edges indicate cases with $Ra < \widetilde{Ra}_{mac}^{\infty}$; filled circles with gold edges indicate cases with $Ra > \widetilde{Ra}_{s}^{\infty}$. The grey diamonds denote RBC ($Ek = \infty$, $Ch = 0$) data from \cite{Vogt2018}; the grey triangles mark rotating convection data from \cite{Vogt_Horn_Aurnou_2021}; hollow squares ($\Lambda=0$) and blue filled squares ($\Lambda=0.72$) denote DNS data from \citet{HornAurnou2025} at $Ek=1.2\times10^{-4}$ with aspect ratio $\Gamma=8$.}
\label{fig:DF_scaling}
\end{figure}

In rotating convection, the ultimate regime—also referred to as the diffusivity-free (DF) regime—describes the limit in which inertia, buoyancy, and Coriolis forces dominate the flow dynamics, while viscous and thermal diffusivities become negligible \citep{JulienKeith2012, Aurnou2020}. The DF regime likely best describes rotating convective turbulence in $\Lambda \ll 1$ planetary and stellar fluid systems \citep{aurnou2015rotating}. Evidence supporting this DF scaling has emerged in recent studies \citep{Lepot_2018, Hadjerci_Bouillaut_Miquel_Gallet_2024, abbate2024, abbate2024diffusivityfree}. For instance, \citet{abbate2024} and \citet{abbate2024diffusivityfree} demonstrate that the heat transfer, internal temperature fluctuations, and flow velocities associated with the thermal-inertial oscillatory mode in liquid metal convection are consistent with DF scaling predictions.

We test our RMC data against the DF scaling to determine whether the flow is dominated by inertial or Lorentz forces. The convective heat transfer, $Nu - 1$, is plotted as a function of the DF scaling parameter $C_JRo_c^2Pe_{ff}$ in Figure~\ref{fig:DF_scaling}a, where the pre-factor $C_J = 1/25$ is taken from \cite{JulienKeith2012}. Both RC and RMC data generally follow the DF scaling (solid black line), with only the data at $Ek = 10^{-4}$ deviating below this trend \citep{abbate2024diffusivityfree}.  
The RMC data at $Ek = 10^{-5}$ exhibit a similar slope to the DF scaling but are shifted upward, which is due to the higher heat transfer efficiency found in these magnetostrophic convection cases.

Figure~\ref{fig:DF_scaling}b plots the vertical Reynolds number $Re_z$ as a function of the DF scaling prediction $C_J^{2/5}Ro_cRe_{ff}$ for RC and RMC cases.
The velocity data for RC (hollow circles and triangles) align well with the DF scaling. In contrast, the RMC data (filled circles) exhibit a transition around $Re_z \approx 800$. Below this threshold, the velocities are suppressed by the magnetic field and fall below the DF scaling prediction. Above this threshold, the inertial effect is still dominant and the RMC data also follow the DF scaling. Notably, the suppression of convective velocity corresponds with enhanced convective heat transfer, arising from increased coherence of the convective flow (see Figure~\ref{fig:velocity_diagram}c,d), a mechanism also observed in turbulent Rayleigh–Bénard convection \citep{Huang2013,Chen_Xie_Yang_Ni_2023}.

Figure \ref{fig:DF_scaling}c shows the vertical Reynolds number normalized by the free-fall Reynolds number $Re_z/Re_{ff}$ plotted against the convective Rossby number $Ro_c$. Following \citet{abbate2024diffusivityfree}, the inclined and horizontal black solid lines, respectively, represent the DF scaling $C_J^{2/5}Ro_c$ for rotating convection (RC) and 
$C_J^{2/5}$ for Rayleigh–Bénard convection (RBC). The RC data, shown as hollow circles and triangles, follow the DF scaling. In contrast, the RMC data fall below the DF scaling but gradually approach it as $Ro_c$ increases. 

Unfortunately, no low $Pr$, slowly rotating ($Ro_c > 1$) RC data exists to plot on the right side on Figure \ref{fig:DF_scaling}c. 
In place of this missing data, the non-rotating, non-magnetic liquid gallium RBC data from \cite{Vogt2018} are included in this plot as hollow grey diamonds. 
The $Ek=\infty$ data, plotted on the right edge of the figure, are well aligned with the non-rotating diffusivity-free scaling prediction using the \citet{JulienKeith2012} prefactor $C_J^{2/5}\approx0.276$.

Figure \ref{fig:DF_scaling}c also includes \citet{HornAurnou2025}'s $\Gamma=8$ cylindrical direct numerical simulation (DNS) data at
$Ek=1.2\times10^{-4}$ (hollow square (RC $\Lambda=0$ data); blue-filled square (RMC $\Lambda=0.72$ data)). Their $\Lambda=0.72$ RMC simulations exhibit lower Reynolds numbers compared to RC cases, consistent with our experimental observations. Their DNS data with $Ro_c>1$ align with the predicted slowly rotating DF scaling. 

Overall, the rapidly rotating ($Ro_c < 1$) RC data in Figure \ref{fig:DF_scaling}c align well with the rapidly rotating DF predictions, $Re \simeq C_J^{2/5} Ro_c Re_{ff}$ \citep{abbate2024diffusivityfree}.  Further, in the slowly (or non-rotating) $Ro_c \gg 1$ regime, we find that $Re \simeq C_J^{2/5} Re_{ff}$. The data in the slowly rotating regime ($Ro_c \gtrsim 1$) at high-$Pr$ fluids are available in \cite{AbbateAurnou2023,abbate2024shadowgraph}, whereas data for low-$Pr$ fluids are currently lacking, and future experiments are needed to confirm the asymptotic slowly rotating scaling. Since all the RMC data lie below the RC data, this implies that the asymptotic, DF RC scalings provide good first-order bounding estimates for  convective velocities in planetary-core and subsurface ocean environments.



\section{Planetary applications}
\label{sec:Application}

\begin{figure}[t!]
\centerline{\includegraphics[width=10.5cm]{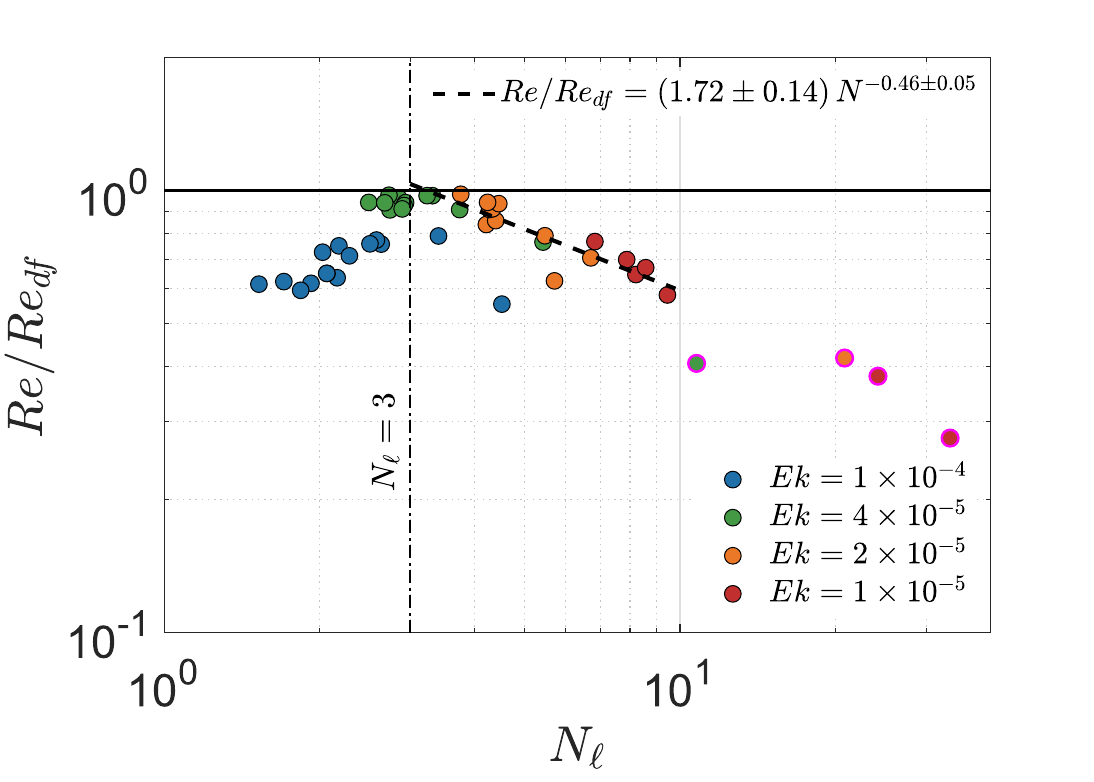}}
  \caption{Reynolds number  normalized by the diffusion-free RC scaling prediction, $Re/Re_{df}$, plotted as a function of local interaction parameter $N_{\ell}$ for RMC data, black dashed line is the best fit the data with $N_{\ell} \leq 3$ (data at $Ek=1\times10^4$ and magnetostrophic stable data (pink edge color) are not included), vertical black dot-dashed line is $N_{\ell}=3$, and horizontal black solid line represents $Re_z=Re_{df}$.} 
\label{fig:Re_div_Redf_fit}
\end{figure}

To further investigate the $N_\ell \approx 3$ transition between geostrophic and magnetically-damped convection, Figure \ref{fig:Re_div_Redf_fit} plots the $\Lambda = 1$ $Re_z$ data normalized by the diffusivity-free RC scaling prediction, $Re/Re_{df}$ as a function of $N_{\ell}$. While the Reynolds number is denoted as $Re_z$ in prior experiments to indicate its vertical component, we generalize the notation to $Re$ in this section. We use $Re_{df}$ as the normalization here (instead of $Re_{cia}$ as used in Figure \ref{fig:Supercriticality_N}) because $Re_{df}$ better extrapolates to planetary settings, where diffusivity-free turbulence is argued to dominate bulk convective flows \citep{stevenson1979, julien2012statistical, barker2014, Bouillaut2021, abbate2024diffusivityfree, vankan2025}.

The magnetically-damped, $N_\ell \geq 3$ data collapse moderately well in Figure \ref{fig:Re_div_Redf_fit}.
%
We therefore perform a fit to the $N_{\ell} \geq 3$ data, excluding one outlier from the $Ek = 1 \times 10^{-4}$ dataset and the magenta-edged points in which the magnetostrophic convective mode is not active.
We refer to the resulting scaling as the magnetically damped Reynolds number, $Re_{MD}$, which is fit as follows:
\begin{equation}
Re_{MD}=cRe_{df}N_{\ell}^{\alpha},
\label{eq:Re_MD_emperical}
\end{equation}
The best fit to \eqref{eq:Re_MD_emperical} is shown as the black dotted line in Figure \ref{fig:Re_div_Redf_fit}, with corresponding best fit parameter values of $c = 1.72 \pm 0.14 \approx 2$ and $\alpha = - 0.46 \pm 0.05 \approx -1/2$. The data that do not have active magnetostrophic modes (magenta edge color) are not included in the above fit.  However, even when these points are included, the best fit exponent remains near -1/2, with $\alpha = -0.53\pm 0.03$.

Thus, 
the empirical best fit for $Re_{MD} / Re_{df}$ 
is
\begin{equation}
\frac{Re_{MD}}{Re_{df}}\approx\frac{2}{N_{\ell}^{1/2}}, 
\quad \text{for } N_{\ell} \gtrsim 3,
\label{eq:Re_MD_fit}
\end{equation}
noting that this relationship holds only in the quasi-static, low local magnetic Reynolds number limit that applies to our laboratory liquid metal MHD experiments. 
This $Re \sim N_\ell^{-1/2} \propto |B|^{-1}$ relation shows that the convective flow velocities are reduced with increasing Lorentz forces (i.e., higher $N_{\ell}$) from the axially imposed magnetic field \citep[cf.][]{vogt2021free}.

We recast \eqref{eq:Re_MD_fit} into a piecewise, regime-dependent form to provide a unified prediction of the Reynolds number across both weakly and strongly magnetically influenced regimes:
\begin{equation}
Re \approx  
\left\{
\begin{array}{ll}
\quad \,\,\, Re_{df} & \text{  for  }   N_{\ell}\lesssim3 \\
2 N_{\ell}^{-1/2} \, Re_{df}  &  \text{  for  }  N_{\ell} \gtrsim3 
\end{array}
\right.
\label{eq:Re_Nc_regime}
\end{equation}
%
%
This form offers a practical framework for estimating planetary core convective velocities based on the local interaction parameter, and highlights the critical threshold $N_{\ell} \approx3$  where magnetic damping becomes pronounced.


Alternatively, it is possible to formulate an equivalent scaling expression for $Re_{MD}$ in terms of input parameters.
This is accomplished by substituting \eqref{eq:N_ell} into \eqref{eq:Re_MD_emperical}, and then by approximating the characteristic length scale to be $\ell = Ro_cH$, which holds in and near the geostrophic turbulence regime \citep{Yadav_2016_PNAS, aurnou2017cross}. 
By doing so, we obtain:
\begin{eqnarray}
Re_{MD} &=& cRe_{df}\left(\frac{\sigma B^2\ell}{\rho U}\right)^{\alpha} \nonumber \\
&=& cRe_{df}\left(\frac{\sigma B^2 (Ro_cH)}{\rho U}\right)^{\alpha} \nonumber \\
&=& c(C_J^{2/5}Ro_cRe_{ff})\left(\frac{ChRo_c}{Re_{MD}}\right)^{\alpha} 
\label{eq:Re_md_derive}
\end{eqnarray}
%
%
%
Recasting \eqref{eq:Re_md_derive} using our best-fit empirical values, $c \approx 2$ and $\alpha \approx -1/2$, then yields:
\begin{equation}
Re_{MD}=4C_J^{4/5}\left(\frac{RaEk}{Pr}\right)^{3/2}\frac{Ek^{1/2}}{\Lambda} = 4C_J^{4/5}  \, Ro_c^{3} \ (Ek \, \Lambda)^{-1}, 
\label{eq:Re_md_N_ell}
\end{equation}
where $4C_J^{4/5} \approx 0.3$. 
Expression \eqref{eq:Re_md_N_ell} captures the $N_\ell \gtrsim 3$ dependence of the magnetically damped Reynolds number solely using input parameters. This enables us to make predictions of convective flow velocities in the magnetostrophic $\Lambda \sim 1$ regime without requiring direct velocity measurements.



The convective interaction parameter $N_c$ and the local interaction parameter $N_{\ell}$ are nearly equivalent in our $N_\ell \gtrsim 3$ data (Figure~\ref{fig:N_ell_vs_N_c}). Either of these parameters can therefore be used to predict the corresponding interaction parameter range in Earth’s outer core. However, $N_c$ is computed directly from the input parameters ($Ra$, $Ek$, $Pr$, and $\Lambda$), whereas $N_{\ell}$ requires estimates of the flow scales and velocity, which are difficult to accurately obtain. We therefore use $N_c$ to estimate the convective interaction parameter range relevant to Earth’s core.

\begin{figure}[t!]
\centerline{\includegraphics[width=10.1cm]{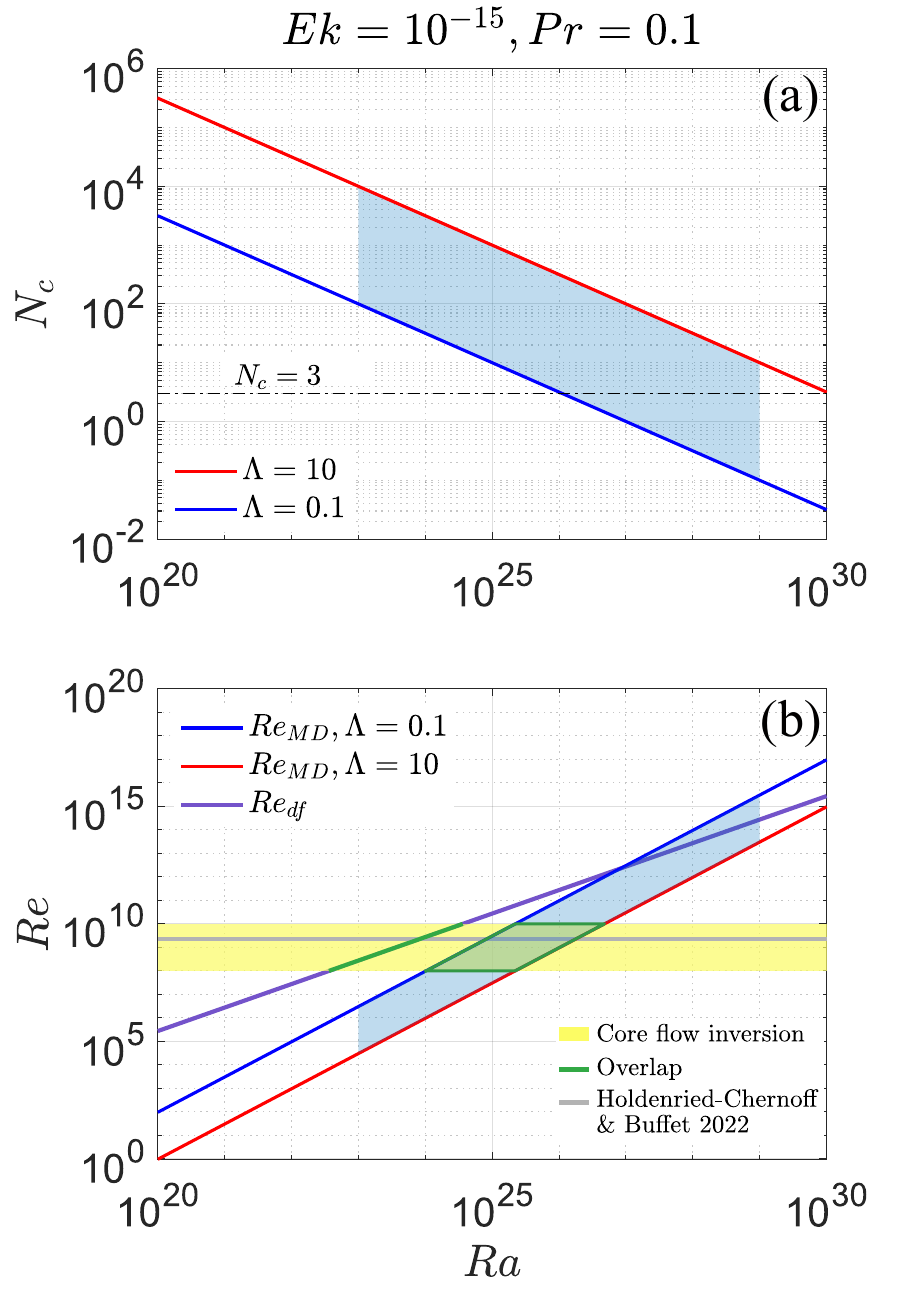}}
  \caption{(a) Convective interaction parameter $N_c$ via \eqref{eq:N_c} and (b) magnetically damped Reynolds number $Re_{MD}$ based on the predictions from \eqref{eq:Re_md_N_ell} as a function Rayleigh number $Ra$,
   using Earth's outer core parameters: $Ek = 10^{-15}$, $Pr = 0.1$. Blue and red solid lines represent $\Lambda = 0.1$ and $\Lambda = 10$, respectively.  The purple solid line marks the DF scaling $Re_{df}$. The semi-transparent blue region shows the estimated range of $Ra$-$\Lambda$ in Earth's outer core corresponding to semi-tranparent rectanlge box shown in Figure \ref{fig:earthcore_linearprediction}b, the light yellow area indicates the corresponding estimated range of $Re$ from \cite{Finlay_2011,Roberts_2013}, the overlap of these estimations is marked by green solid line or green border lozenge, grey solid line from \citet{holdenried2022}.}
\label{fig:Re_MD_vs_Ra}
\end{figure}

Figure \ref{fig:Re_MD_vs_Ra}a shows predictions of $N_c$ for the two Elsasser number limits, $\Lambda = 0.1$ and $\Lambda = 10$ based on Earth's core estimates of $Ek = 10^{-15}$ and $Pr = 0.1$, plotted as blue and red solid lines, respectively. The semi-transparent blue shading denotes the estimated $N_c$–$Ra$ range for Earth’s core, while the black dash-dotted line marks the transition threshold $N_c = 3$. Most estimates fall above this threshold, placing Earth’s core in the magnetically damped regime, consistent with our experimental results.

Figure \ref{fig:Re_MD_vs_Ra}b presents $Re_{MD}$, as expressed via \eqref{eq:Re_md_N_ell}, as a function of the Rayleigh number $Ra$. The predictions are shown for two Elsasser numbers, $\Lambda = 0.1$ (blue solid line) and $\Lambda = 10$ (red solid line), using Earth's core estimates of $Ek = 10^{-15}$ and $Pr = 0.1$. The semi-transparent blue region indicates estimated outer core ranges of $Ra \in [10^{23}, 10^{29}]$ and $\Lambda \in [0.1, 1]$ (identical to the blue region in Figure \ref{fig:earthcore_linearprediction}b). The light yellow rectangle represents the estimated range of $Re$ in Earth's core \citep[e.g.,][]{Finlay_2011, Holme2015, livermore2017}.
%
The overlap of the blue and yellow shaded regions, edged in green, defines the range of $Ra$ and $Re_{MD}$ consistent with our laboratory experimental predictions. 
The center of the overlap region in Figure \ref{fig:Re_MD_vs_Ra} corresponds to 
an Earth's core Rayleigh number estimate of $Ra \approx 5 \times 10^{25}$, with the edges of the overlap region corresponding to Rayleigh values ranging from $Ra \approx 10^{24}$ to $5 \times 10^{26}$. These values are in good agreement with previous estimates \cite[e.g.,][]{Gubbins2001}.

The purple solid line in Figure \ref{fig:Re_MD_vs_Ra}b corresponds to $Re_{df}$ scaling \eqref{eq:ReDF}, which is close to $Re_{MD}=(2/\sqrt{3})Re_{df}$ \eqref{eq:Re_Nc_regime} at $N_{\ell}=3$ according to the best-fit obtained in Figure~\ref{fig:Re_div_Redf_fit}.  
This diffusion-free RC model for the convective flow velocities intersects the yellow shaded region between $Ra \approx 3 \times 10^{22}$ and $3 \times 10^{24}$. These DF branch predictions are still within the range of estimated core $Ra$, and, according to our extrapolations, likely constitute lower bounding $Ra$ estimates for convection in Earth's outer core. The Earth’s core–like range (semi-transparent blue region) lies mostly below $Re_{df}$, indicating a magnetically damped flow regime. This is consistent with Figure \ref{fig:Re_MD_vs_Ra}a, which shows that the predicted convective interaction parameter $N_c$ for the Earth’s core remains largely above $N_c>3$, where the flow is magnetically damped.

In making these Figure \ref{fig:Re_MD_vs_Ra} estimates for $Ra$ in Earth's core, it must be pointed out that we are using large-scale core-mantle boundary flow inversion results as constraints on the small-scale convective flow within the bulk core.  This may indeed be accurate and appropriate, but it is possible that the large-scale inferred surface velocities differ substantively from local convective-scale flows deep within the core \citep[e.g.,][]{Holme2015, holdenried2022, oliver2023}.  To address this point, we include the grey horizontal line that denotes the turbulent diffusivity estimate from \citet{holdenried2022} for bulk rms outer core velocity. This estimate agrees well with that of the velocity estimate from core flow inversion, and thus supports our Figure \ref{fig:Re_MD_vs_Ra} predictions.

\section{Conclusion}
\label{Conclu}

We have presented the results of a series of laboratory experiments in rotating magnetoconvection using liquid gallium to systematically explore the parameter regime characterized by moderate Ekman numbers and Elsasser numbers of order unity.  Rotating convection cases were also investigated to provide a baseline for comparison with the RMC data. This study presents the first direct measurements of convective velocities in the $\Lambda = 1$, $Pr \ll 1$ magnetostrophic regime, a regime theoretically predicted to govern the local convection dynamics in planetary core settings \citep{aurnou2017cross, Teed2025}.

At $Ek = 10^{-4}$, the flow is not strongly influenced by the presence of an imposed $\Lambda = 1$ magnetic field. This is because inertial effects are equivalent to, or dominate, over both Coriolis and Lorentz forces in these $Ek = 10^{-4}$ experiments. In contrast, at $Ek = 10^{-5}$, inertial accelerations are subdominant, allowing the magnetic field to significantly alter the flow pattern, leading to larger-scale structures and reduced turbulence intensity. 


Quantitatively, we find convective heat transfer and flow velocities in RMC cases exhibit a relatively sharp transition at a local interaction parameter $N_{\ell}\approx3$. Above this threshold, the heat transfer is enhanced, which has been previously reported in both liquid metal experimental studies \citep{KingAurnou2015PNAS, Grannan_2022}, but has been interpreted to be due to the presence of magneto-wall modes or magnetostrophic bulk modes based only on sparse thermal field measurements. Here,  we directly measure the suppression of bulk convective velocities in the $N_\ell \gtrsim 3$ regime is reported for the first time in liquid metal RMC experiments. These suppressed magnetostrophic convective velocities are accompanied by enhanced convective heat transfer. This enhancement may be explained by an increased coherence of the magneto-suppressed flow field, an argument that is qualitatively similar to that found in the constrained Rayleigh–Bénard convection literature \citep{Huang2013, Chen_Xie_Yang_Ni_2023}. Below the $N_\ell \approx 3$ threshold, the measured velocities consistently scale with both the CIA and the diffusivity-free RC predictions, indicating that the flow is in the  geostrophic turbulent regime. 
In addition, the $N_\ell \approx 3$ transition found in our quasi-static liquid metal experiments is in basic agreement with the $N_\ell \approx 2$ dipolarity transition found in the spherical shell dynamo modeling study of \citet{Soderlund2025}.



We derived a set of predictive formulations for the Reynolds number, \eqref{eq:Re_Nc_regime} and \eqref{eq:Re_md_N_ell}, that depend solely on externally estimable parameters ($Ra$, $Ek$, $Pr$, $\Lambda$), enabling core velocity estimations to be made without direct measurements. 
Applying these experimental velocity scalings to estimate the convective Reynolds number in Earth’s outer core, our $Re_{MD}$-based predictions align well with previous geophysical estimates and suggest that convection in Earth's outer core occurs at a Rayleigh number of $Ra \approx 5 \times 10^{25}$ and a convective Rossby number of $Ro_c \approx 2 \times 10^{-2}$, parameter values which are notoriously difficult to obtain for deep planetary core interiors.

Our experimental results also have implications for planetary dynamo models. Most $Ek \gtrsim 10^{-5}$ dynamo models are in the `weak field' regime in which Lorentz forces are significantly smaller than Coriolis \citep[e.g.,][]{schaeffer2017}.  Even at the small, local scale of hydrodynamic RC onset, the local Elsasser number is typically below unity in these high $Ek$ models \citep[][]{aurnou2017cross}.  Our results imply that the velocities in such models should tend towards the diffusivity-free RC scaling, as this provides the upper bounding velocities when $N_\ell \lesssim 1$, as shown in Figure \ref{fig:DF_scaling}c. In contrast, when $Ek \lesssim 10^{-5}$, `strong field' dynamo action tends to occur in which the local scale Elsasser number is order one or greater and the magnetic energy tends to swamp the kinetic energy \citep{Yadav_2016_PNAS, schaeffer2017, dormy2025}.  In these models, we expect locally magnetostrophic convective flows to be larger-scale and magnetically damped in their intensity, as found here in our $N_\ell \gtrsim 3$ cases. Since the vast majority of dynamo simulations to date have been carried out with $Ek \gtrsim 10^{-5}$, we predict that most will tend to track well with diffusivity-free rotating convective velocity scaling predictions \cite[e.g.,][]{Aurnou2020, Hadjerci_Bouillaut_Miquel_Gallet_2024, abbate2024diffusivityfree}. 

\section{Acknowledgement}
We gratefully acknowledge funding for this project from the NSF Geophysics Program via EAR award \#2143939.

\appendix
\section{Appendix}
\label{sec:appendix}

\begin{figure}[htbp!]
  \centerline{\includegraphics[width=14cm]{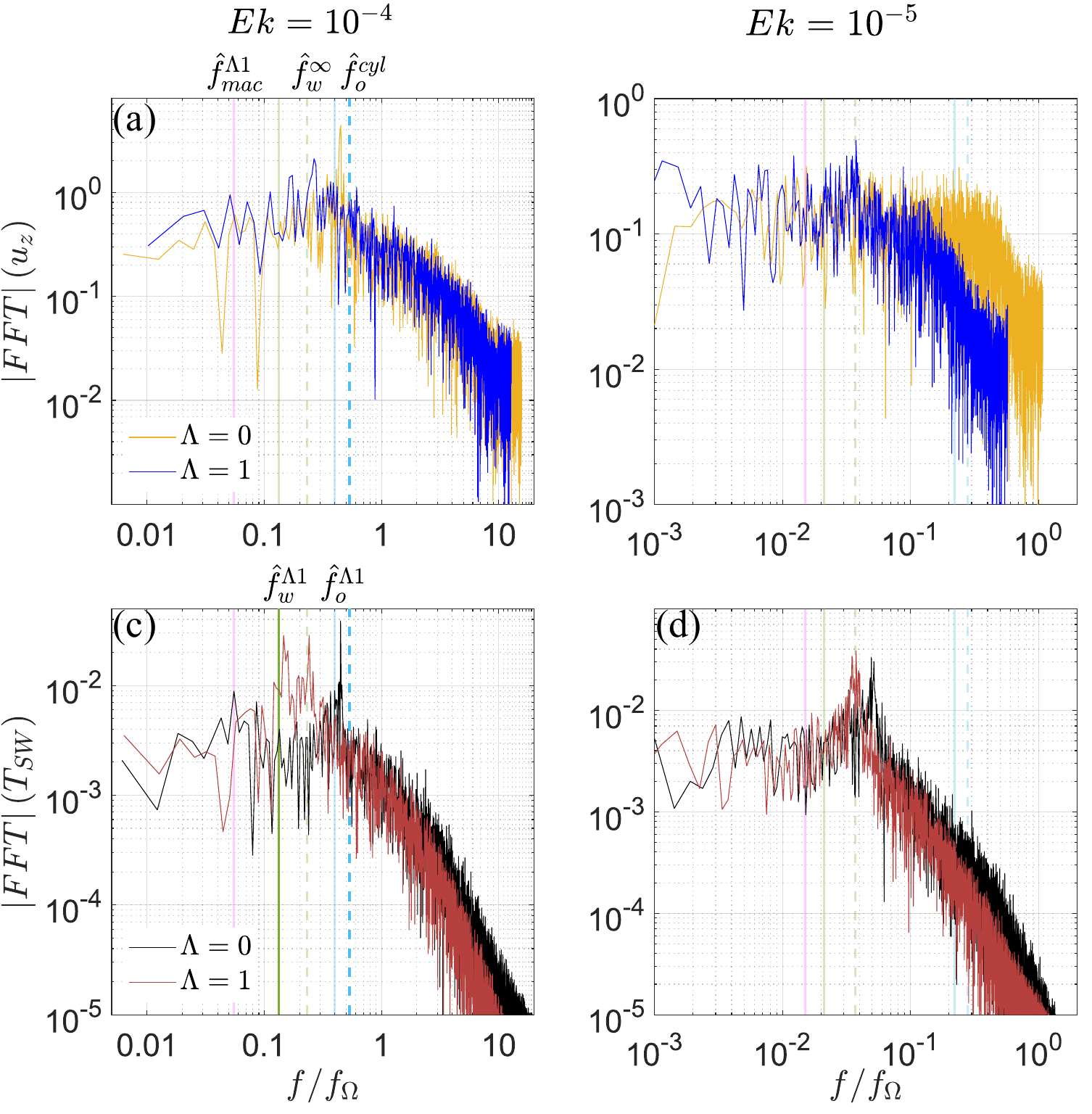}}
  \caption{Amplitude of the Fourier Transform of vertical velocity $u_z$: gold  $\Lambda=0$ and blue $\Lambda=1$ and of side-wall temperature signal: black $\Lambda=0$ and brick red $\Lambda=1$ ($a, c$) $Ek=10^{-4}$ and ($b,d$) $Ek=10^{-5}$; the velocities data correspond to Figure~\ref{fig:velocity_diagram}; pink dashed line: mac frequency prediction $\hat{f}_{mac}^{\Lambda1}$; green solid line: wall mode frequency prediction $\hat{f}_{w}^{\Lambda1}$ with $\Lambda=1$; green dashed line: wall mode frequency prediction $\hat{f}_{w}^{\infty}$ with $\Lambda=0$;
  light blue solid line: oscillatory mode frequency prediction $\hat{f}_{o}^{\Lambda1}$ with $\Lambda=1$; light blue dashed line: oscillatory mode frequency prediction $\hat{f}_{o}^{cyl}$ with $\Lambda=0$.}
\label{fig:FFT_velocity_SWtemperature_comparison}
\end{figure}



\newpage
\bibliographystyle{plainnat}
\bibliography{Manuscript_bib}

\appendix
\section{Appendix}
\begin{landscape}
\setlength{\tabcolsep}{4pt}      
\renewcommand{\arraystretch}{0.9}
\begin{table}[H]
  \begin{center}
\def~{\hphantom{0}}
  \begin{tabular}{cccccccccccccc}
          $\Lambda$ & $Ek\cdot{10^{4}}$ & $Ra\cdot{10^{-6}}$ &  $\widetilde{Ra}_o^{\infty}$  & $\widetilde{Ra}_o^{cyl}$ &$\widetilde{Ra}_w^{\infty}$ &$\widetilde{Ra}_{mac}^{\infty}$ & $\widetilde{Ra}_{s}^{\infty}$&  $Pr$ & P(W) & $\Delta T$ & $Nu$ & $Re_{z,rms}$  & $Ro_c$ \\[3pt]
          \hline
          0 & 1 & 0.335 & 10.55 & 3.39 & 1.40 & $-$ & 0.18 & 0.0275 & 20 & 1.33 & 1.44 & 265 & 0.345\\
          0 & 1 & 0.615 & 19.37 & 6.22 & 2.57 & $-$ & 0.32 & 0.0275 & 50 & 2.44 & 2.03 & 534 & 0.467\\
          0 & 1 & 0.783 & 24.67 & 7.92 & 3.27 & $-$ & 0.41 & 0.0275 & 80 & 3.08 & 2.57 & 911 & 0.525\\
          0 & 1 & 3.08 &  97.03  & 31.15 & 12.88 & $-$ & 1.62 & 0.0275 & 780 & 12.01 & 6.69 & 2566 & 1.04\\
          1 & 1 & 0.329 & 1.17 & $-$ & 1.41 & 0.58 & 0.19 & 0.0275 & 20 & 1.31 & 1.48 & $-$ & 0.342\\
          1 & 1 & 0.600 & 2.14 & $-$ & 2.58 & 1.05 & 0.34 & 0.0275 & 50 & 2.37 & 2.09 & $-$ & 0.461\\ 
          1 & 1 & 0.785 & 2.80 & $-$ & 3.37 & 1.37 & 0.44 & 0.0272 & 80 & 3.08 & 2.62 & 326 & 0.526\\ 
          1 & 1 & 1.17 & 4.17 & $-$ & 5.02 & 2.05 & 0.66 & 0.0273 & 150 & 4.61 & 3.3 & 603 & 0.643\\ 
          1 & 1 & 1.41 & 5.02 & $-$ & 6.05 & 2.47 & 0.79 & 0.0273 & 200 & 5.56 & 3.69 & 864 & 0.706\\ 
          1 & 1 & 1.63 & 5.81 & $-$ & 7.00 & 2.85 & 0.92 & 0.0272 & 250 & 6.40 & 4.02 & 1065 & 0.758\\ 
          1 & 1 & 1.85 & 6.59 & $-$ & 7.94 & 3.24 & 1.04 & 0.0271 & 300 & 7.21 & 4.3  & 1202 & 0.804\\ 
          1 & 1 & 2.04 & 7.27 & $-$ & 8.76 & 3.57 & 1.15 & 0.0270 & 350 & 7.96 & 4.55 & 1340 & 0.845\\ 
          1 & 1 & 2.24 & 7.98 & $-$ & 9.62 & 3.92 & 1.26 & 0.0269 & 400 & 8.69 & 4.74 & 1430 & 0.883\\ 
          1 & 1 & 2.44  & 8.69 & $-$ & 10.48 & 4.27 & 1.37 & 0.0268 & 450 & 9.43 & 4.94 & 1532 & 0.92\\ 
          1 & 1 & 2.62 & 9.33 & $-$ & 11.25 & 4.58 & 1.47 & 0.0267 & 500 & 10.12 & 5.13 & 1484 & 0.963\\ 
          1 & 1 & 2.80 & 9.97 & $-$ & 12.02 & 4.90 & 1.58 & 0.0267 & 550 & 10.77 & 5.28 & 1629 & 0.983\\ 
          1 & 1 & 2.98 & 10.62 & $-$ & 12.79 & 5.21 & 1.68 & 0.0266 & 600 & 11.45 & 5.45 & 1743 & 1.01\\ 
          1 & 1 & 3.13 & 11.15 & $-$ & 13.44 & 5.47 & 1.76 & 0.0265 & 650 & 11.99 & 5.63 & 1849 & 1.04\\ 
          1 & 1 & 3.31 & 11.79 & $-$ & 14.21 & 5.79 & 1.86 & 0.0264 & 700 & 12.65 & 5.76 & 1933 & 1.07\\
          1 & 1 & 3.57 & 12.72 & $-$ & 15.33 & 6.24 & 2.01 & 0.0264 & 780 & 13.61 & 5.96 & 2005 & 1.1\\ 
  \end{tabular}
  \caption{Data table for the $Ek = 10^{-4}$ experiments.}
  \label{table:summary1}
  \end{center}
\end{table}
\end{landscape}

\begin{landscape}
\setlength{\tabcolsep}{4pt}      
\renewcommand{\arraystretch}{0.9}
\begin{table}[H]
  \begin{center}
\def~{\hphantom{0}}
  \begin{tabular}{cccccccccccccc}
          $\Lambda$ & $Ek\cdot{10^{4}}$ & $Ra\cdot{10^{-6}}$ &  $\widetilde{Ra}_o^{\infty}$  & $\widetilde{Ra}_o^{cyl}$ &$\widetilde{Ra}_w^{\infty}$ &$\widetilde{Ra}_{mac}^{\infty}$ & $\widetilde{Ra}_{s}^{\infty}$&  $Pr$ & P(W) & $\Delta T$ & $Nu$ & $Re_{z,rms}$  & $Ro_c$ \\[3pt]
          \hline
          0 & 0.4 & 0.441 & 4.40 & 1.52 & 0.64 & $-$ & 0.07 & 0.0275 & 20 & 1.75 & 1.09 & $-$ & 0.159\\
          0 & 0.4 & 1.49 & 14.87 & 5.15 & 2.15 & $-$ & 0.23 & 0.0272 & 100 & 5.86 & 1.71 & 492 & 0.290\\
          0 & 0.4 & 1.93 & 19.26 & 6.67 & 2.78 & $-$ & 0.30 & 0.0270 & 150 & 7.54 & 2.00 & 835 & 0.330\\
          0 & 0.4 & 2.26 & 22.56 & 7.81 & 3.26 & $-$ & 0.35 & 0.0270 & 200 & 8.82 & 2.31 & 766 & 0.357\\
          0 & 0.4 & 5.02 & 50.10 & 17.34 & 7.24 & $-$ & 0.78 & 0.0261 & 780 & 18.95 & 4.27 & 1786 & 0.523\\
          1 & 0.4 & 0.441 & 0.60 & $-$ & 0.73 & 0.31 & 0.07 & 0.0275 & 20 & 1.75 & 1.09 & 216 & 0.159\\
          1 & 0.4 & 1.36 & 1.84 & $-$ & 2.26 & 0.95 & 0.22 & 0.0272 & 100 & 5.33 & 1.88 & 517 & 0.277\\
          1 & 0.4 & 1.74 & 2.36 & $-$ & 2.89 & 1.22 & 0.28 & 0.0271 & 150 & 6.79 & 2.22 & 726 & 0.313\\
          1 & 0.4 & 2.06 & 2.79 & $-$ & 3.43 & 1.44 & 0.34 & 0.0271 & 200 & 8.05 & 2.54 & 977 & 0.341\\
          1 & 0.4 & 2.59 & 3.51 & $-$ & 4.31 & 1.82 & 0.42 & 0.0269 & 300 & 10.08 & 3.06 & 1072 & 0.381\\
          1 & 0.4 & 2.86 & 3.87 & $-$ & 4.76 & 2.01 & 0.47 & 0.0267 & 350 & 11.05 & 3.26 & 1139 & 0.399\\
          1 & 0.4 & 3.08 & 4.17 & $-$ & 5.12 & 2.16 & 0.50 & 0.0268 & 400 & 11.91 & 3.35 & 1243 & 0.414\\
          1 & 0.4 & 3.33 & 4.51 & $-$ & 5.54 & 2.34 & 0.54 & 0.0266 & 450 & 12.79 & 3.63 & 1274 & 0.429\\
          1 & 0.4 & 3.55 & 4.81 & $-$ & 5.91 & 2.49 & 0.58 & 0.0266 & 500 & 13.62 & 3.80 & 1351 & 0.443\\
          1 & 0.4 & 3.78 & 5.12 & $-$ & 6.29 & 2.65 & 0.62 & 0.0264 & 550 & 14.43 & 3.93 & 1401 & 0.456\\
          1 & 0.4 & 4.00 & 5.42 & $-$ & 6.65 & 2.81 & 0.65 & 0.0263 & 600 & 15.21 & 4.10 & 1429 & 0.468\\
          1 & 0.4 & 4.39 & 5.94 & $-$ & 7.30 & 3.08 & 0.72 & 0.0262 & 700 & 16.63 & 4.37 & 1553 & 0.490\\
          1 & 0.4 & 4.67 & 6.32 & $-$ & 7.77 & 3.27 & 0.76 & 0.0262 & 780 & 17.64 & 4.58 & 1596 & 0.504\\
  \end{tabular}
  \caption{Data table for the $Ek = 4 \times 10^{-5}$ experiments.}
  \label{table:summary1}
  \end{center}
\end{table}
\end{landscape}

\begin{landscape}
\setlength{\tabcolsep}{4pt}      
\renewcommand{\arraystretch}{0.9}
\begin{table}[H]
  \begin{center}
\def~{\hphantom{0}}
  \begin{tabular}{cccccccccccccc}
          $\Lambda$ & $Ek\cdot{10^{4}}$ & $Ra\cdot{10^{-6}}$ &  $\widetilde{Ra}_o^{\infty}$  & $\widetilde{Ra}_o^{cyl}$ &$\widetilde{Ra}_w^{\infty}$ &$\widetilde{Ra}_{mac}^{\infty}$ & $\widetilde{Ra}_{s}^{\infty}$&  $Pr$ & P(W) & $\Delta T$ & $Nu$ & $Re_{z,rms}$  & $Ro_c$ \\[3pt]
          \hline
          0 & 0.2 & 0.949 & 3.88 & 1.43 & 0.64 & $-$ & 0.06 & 0.0273 & 40 & 3.74 & 1.03 & 325 & 0.124\\
          0 & 0.2 & 1.97 & 8.05 & 2.97 & 1.33 & $-$ & 0.12 & 0.0271 & 100 & 7.71 & 1.29 & 589 & 0.177\\
          0 & 0.2 & 2.63 & 10.74 & 3.97 & 1.77 & $-$ & 0.16 & 0.0269 & 150 & 10.21 & 1.46 & 746 & 0.204\\
          0 & 0.2 & 6.89 & 28.15 & 10.40 & 4.65 & $-$ & 0.43 & 0.0259 & 780 & 25.85 & 3.11 & 1580 & 0.305\\
          1 & 0.2 & 0.968 & 0.63 & $-$ & 0.79 & 0.34 & 0.06 & 0.0273 & 40 & 3.81 & 1.01 & $-$ & 0.125\\
          1 & 0.2 & 1.73 & 1.12 & $-$ & 1.42 & 0.60 & 0.11 & 0.0271 & 100 & 6.76 & 1.48 & $-$ & 0.166\\
          1 & 0.2 & 2.32 & 1.51 & $-$ & 1.90 & 0.81 & 0.15 & 0.0270 & 150 & 9.01 & 1.67 & $-$ & 0.192\\
          1 & 0.2 & 2.83 & 1.84 & $-$ & 2.32 & 0.99 & 0.18 & 0.0267 & 200 & 10.91 & 1.86 & 212 & 0.211\\
          1 & 0.2 & 3.33 & 2.16 & $-$ & 2.72 & 1.16 & 0.21 & 0.0266 & 250 & 12.66 & 2.01 & 446 & 0.227\\
          1 & 0.2 & 3.68 & 2.39 & $-$ & 3.01 & 1.28 & 0.24 & 0.0266 & 300 & 14.12 & 2.18 & 561 & 0.240\\
          1 & 0.2 & 4.03 & 2.62 & $-$ & 3.30 & 1.40 & 0.26 & 0.0266 & 350 & 15.47 & 2.32 & 690 & 0.251\\
          1 & 0.2 & 4.31 & 2.80 & $-$ & 3.53 & 1.50 & 0.28 & 0.0266 & 400 & 16.56 & 2.47 & 783 & 0.260\\
          1 & 0.2 & 4.66 & 3.03 & $-$ & 3.81 & 1.62 & 0.30 & 0.0266 & 450 & 17.91 & 2.58 & 864 & 0.270\\
          1 & 0.2 & 4.95 & 3.22 & $-$ & 4.05 & 1.73 & 0.32 & 0.0265 & 500 & 18.94 & 2.73 & 969 & 0.278\\
          1 & 0.2 & 5.17 & 3.36 & $-$ & 4.23 & 1.80 & 0.33 & 0.0264 & 550 & 19.75 & 2.86 & 1014 & 0.284\\
          1 & 0.2 & 5.53 & 3.59 & $-$ & 4.52 & 1.93 & 0.36 & 0.0264 & 600 & 21.11 & 2.95 & 1113 & 0.294\\
          1 & 0.2 & 5.86 & 3.81 & $-$ & 4.79 & 2.04 & 0.38 & 0.0264 & 680 & 22.33 & 3.14 & 1185 & 0.302\\
          1 & 0.2 & 6.32 & 4.11 & $-$ & 5.17 & 2.20 & 0.41 & 0.0262 & 780 & 23.98 & 3.36 & 1327 & 0.313\\
  \end{tabular}
  \caption{Data table for the $Ek = 2 \times 10^{-5}$ experiments.}
  \label{table:summary1}
  \end{center}
\end{table}
\end{landscape}

\begin{landscape}
\setlength{\tabcolsep}{4pt}      
\renewcommand{\arraystretch}{0.9}
\begin{table}[H]
  \begin{center}
\def~{\hphantom{0}}
  \begin{tabular}{cccccccccccccc}
          $\Lambda$ & $Ek\cdot{10^{4}}$ & $Ra\cdot{10^{-6}}$ &  $\widetilde{Ra}_o^{\infty}$  & $\widetilde{Ra}_o^{cyl}$ &$\widetilde{Ra}_w^{\infty}$ &$\widetilde{Ra}_{mac}^{\infty}$ & $\widetilde{Ra}_{s}^{\infty}$&  $Pr$ & P(W) & $\Delta T$ & $Nu$ & $Re_{z,rms}$  & $Ro_c$ \\[3pt]
          \hline
          0 & 0.1 & 1.23 & 2.03 & 0.80 & 0.40 & $-$ & 0.03 & 0.0272 & 50 & 4.81 & 1.01 & $-$ & 0.069\\
          0 & 0.1 & 2.52 & 4.17 & 1.64 & 0.82 & $-$ & 0.06 & 0.0268 & 100 & 9.76 & 1.07 & $-$ & 0.099\\
          0 & 0.1 & 4.17 & 6.89 & 2.71 & 1.35 & $-$ & 0.10 & 0.0267 & 200 & 16.01 & 1.25 & 289 & 0.127\\
          0 & 0.1 & 5.73 & 9.47 & 3.73 & 1.86 & $-$ & 0.14 & 0.0263 & 300 & 21.76 & 1.40 & 530 & 0.148\\
          0 & 0.1 & 10.10 & 16.69 & 6.57 & 3.28 & $-$ & 0.25 & 0.0253 & 780 & 36.89 & 2.17 & 1289 & 0.192\\
          1 & 0.1 & 1.25 & 0.38 & $-$ & 0.51 & 0.22 & 0.03 & 0.0272 & 50 & 4.88 & 1.00 & $-$ & 0.070\\
          1 & 0.1 & 2.38 & 0.73 & $-$ & 0.97 & 0.42 & 0.06 & 0.0268 & 100 & 9.21 & 1.07 & $-$ & 0.096\\
          1 & 0.1 & 3.59 & 1.10 & $-$ & 1.46 & 0.63 & 0.09 & 0.0267 & 200 & 13.87 & 1.46 & 106 & 0.118\\
          1 & 0.1 & 4.84 & 1.48 & $-$ & 1.97 & 0.85 & 0.12 & 0.0265 & 300 & 18.49 & 1.65 & 194 & 0.136\\
          1 & 0.1 & 7.02 & 2.14 & $-$ & 2.85 & 1.23 & 0.18 & 0.0260 & 550 & 26.43 & 2.13 & 424 & 0.163\\
          1 & 0.1 & 7.49 & 2.29 & $-$ & 3.05 & 1.31 & 0.19 & 0.0259 & 600 & 28.03 & 2.20 & 501 & 0.168\\
          1 & 0.1 & 7.86 & 2.40 & $-$ & 3.20 & 1.37 & 0.20 & 0.0258 & 650 & 29.31 & 2.28 & 542 & 0.171\\
          1 & 0.1 & 8.23 & 2.51 & $-$ & 3.35 & 1.44 & 0.21 & 0.0257 & 700 & 30.57 & 2.36 & 590 & 0.175\\
          1 & 0.1 & 8.80 & 2.69 & $-$ & 3.58 & 1.54 & 0.22 & 0.0256 & 780 & 32.57 & 2.47 & 691 & 0.181\\
  \end{tabular}
  \caption{Data table for the $Ek = 10^{-5}$ experiments.}
  \label{table:summary1}
  \end{center}
\end{table}
\end{landscape}

\end{document}